\documentclass[sigconf]{acmart}

\usepackage{multirow}
\usepackage{graphicx}
\usepackage{caption}
\usepackage{subcaption}
\usepackage{enumitem}
\usepackage{balance}

\copyrightyear{2020}
\acmYear{2020}
\setcopyright{acmcopyright}\acmConference[SIGIR '20]{Proceedings of the 43rd International ACM SIGIR Conference on Research and Development in Information Retrieval}{July 25--30, 2020}{Virtual Event, China}
\acmBooktitle{Proceedings of the 43rd International ACM SIGIR Conference on Research and Development in Information Retrieval (SIGIR '20), July 25--30, 2020, Virtual Event, China}
\acmPrice{15.00}
\acmDOI{10.1145/3397271.3401166}
\acmISBN{978-1-4503-8016-4/20/07}

\begin{document}
\fancyhead{}

\title{Answer Ranking for Product-Related Questions via Multiple Semantic Relations Modeling}

\author{Wenxuan Zhang, Yang Deng, Wai Lam}
\affiliation{%
  \institution{The Chinese University of Hong Kong}
}
\email{{wxzhang, ydeng, wlam}@se.cuhk.edu.hk}

\begin{abstract}
  Many E-commerce sites now offer product-specific question answering platforms for users to communicate with each other by posting and answering questions during online shopping. However, the multiple answers provided by ordinary users usually vary diversely in their qualities and thus need to be appropriately ranked for each question to improve user satisfaction. It can be observed that product reviews usually provide useful information for a given question, and thus can assist the ranking process. In this paper, we investigate the answer ranking problem for product-related questions, with the relevant reviews treated as auxiliary information that can be exploited for facilitating the ranking. We propose an answer ranking model named MUSE which carefully models multiple semantic relations among the question, answers, and relevant reviews. Specifically, MUSE constructs a multi-semantic relation graph with the question, each answer, and each review snippet as nodes. Then a customized graph convolutional neural network is designed for explicitly modeling the semantic relevance between the question and answers, the content consistency among answers, and the textual entailment between answers and reviews. Extensive experiments on real-world E-commerce datasets across three product categories show that our proposed model achieves superior performance on the concerned answer ranking task. 
\end{abstract}
\thanks{The work described in this paper is substantially supported by grants from the Research Grant Council of the Hong Kong Special Administrative Region, China (Project Code: 14204418) and the Direct Grant of the Faculty of Engineering, CUHK (Project Code: 4055093).}

\maketitle

\section{Introduction}
\begin{table}
  \small
  \setlength{\abovecaptionskip}{2pt}   
  \caption{An example question of a headphone product from \textit{Amazon} accompanied with its multiple answers. There are also some relevant review snippets of the question.}
  \label{intro-example}
  \begin{tabular}{p{8.2cm}}
    \toprule
    \textbf{Question}: \; Will these work with Android? \\ 
    \midrule
    \textbf{Relevant Review Snippets:} \\
    - I have a Samsung Galaxy Note 4, I absolutely love these headphones, totally worth what I paid for them. \\
    - Easy to sync with an iPhone/Android phones. \\
    - Waste of money if using with an android device \\
    - I use mine with an android device and still works great. \\
    - ...... \\
    \midrule
    \textbf{Answers:} \\
    A1: The will and despite what some had said, they will sound exactly as they do on an Apple device. \\
    A2: Yes they will work with any Bluetooth capable device. \\
    A3: They will work but they won't sound as proficient as it would with an apple product. \\
    A4: I have Samsung Note 5, the headphones cannot connect via Bluetooth. \\
    A5: Its really not worth the money when using with an android. \\
 \bottomrule
    \end{tabular}
\vspace{-0.3cm}
\end{table}

Appropriately addressing users' concerns during online shopping can greatly improve their shopping experience and stimulate the purchase decisions. To this end, many E-commerce sites such as Amazon and eBay now offer product-specific community question answering platforms for users to post their questions and answer existing questions. Thanks to the convenience of such platforms, a question can typically get multiple user-provided answers. However, these answers, similar with other user-generated content, vary a lot in their qualities \cite{DBLP:conf/wsdm/answer-unreliable, DBLP:conf/sigir/answer-unreliable-2} and suffer from typical flaws such as spams or even malicious content from the competitors. 

Table \ref{intro-example} presents an example question, as well as its user-provided answers ranked by the community votes. It can be observed that even for this relatively objective question, the answers can vary diversely. Such variation in the answer contents and qualities motivates the task of automatically ranking these answers to improve user satisfaction. As shown in the example, there usually exists some relevant product reviews for the concerned question, which can provide useful information when ranking the answers if they are effectively utilized. Thus in this paper, we aim to tackle the task of answer ranking for product-related questions, with the associated product reviews treated as auxiliary information which can be exploited for assisting the ranking.

Answer selection methods have been extensively studied for tackling the answer ranking problem in retrieval-based question answering (QA) systems~\cite{DBLP:conf/acl/TanSXZ16, DBLP:conf/semeval/NakovHMMMBV17, DBLP:conf/emnlp/WangSM07, DBLP:conf/sigdial/EricKCM17}. Most of the existing works focus on measuring the semantic relevance between the question and a candidate answer, where the negative answers for training are usually randomly sampled from the whole answer pool \cite{DBLP:conf/ijcai/Qiu-yahooqa-15, DBLP:conf/sigir/yitay-yahooqa, aaai20-dy} or chosen from irrelevant documents \cite{DBLP:conf/emnlp/wikiqa}. 
However, as can be observed from the above example, merely measuring the semantic relevance between the question and answer texts is no longer sufficient for the concerned answer ranking task in E-commerce settings since all of the answers are written specifically for the question and thus most of them are supposed to be topically relevant to the question.
Moreover, existing general answer selection models lack the capability of making use of the relevant product reviews during the ranking process. 
Recently, \citet{Zhang-WWW2020-rahp} attempt to utilize reviews for identifying helpful answers in E-commerce. However, they ignore the relations among answers and reviews, which is essential for the concerned answer ranking problem.
On the other hand, some product-related question answering (PQA) methods explore the utilization of product reviews to provide responses for a given question~\cite{DBLP:conf/wsdm/YuL18,DBLP:conf/aaai/0007GZZWZS19, DBLP:conf/wsdm/ChenLJZC19,DBLP:conf/wsdm/GaoRZZYY19}, where some selected reviews can be served directly as the response  \cite{DBLP:conf/emnlp/YuZC12,DBLP:conf/aaai/0007GZZWZS19} or used to generate a response sentence based on a sequence-to-sequence neural model \cite{DBLP:conf/wsdm/ChenLJZC19,DBLP:conf/wsdm/GaoRZZYY19}. These PQA methods assume the lack of user-written answers and thus turn to product reviews for help when addressing the given question. Such assumption neglects the large number of available answers provided by former buyers \cite{DBLP:conf/wsdm/YuQJHSCC18,DBLP:conf/sigir/CarmelLM18}, which can better answer the given question than the responses produced from the reviews.

In E-commerce settings, a key issue for the concerned ranking task is what makes an answer be a good one and how we can characterize such good answers via exploiting the associated reviews. 
Specifically, appropriately ranking the user-provided answers requires to model the complex semantic relations among these existing information sources, i.e., the question, answers, and product reviews.
We argue that three kinds of semantic relations attach great importance:
(i) Firstly, similar to general answer ranking problem \cite{rao2019bridging}, the \textit{semantic relevance} of the answer content to the question is still essential for determining the ranking. At the same time, it is also necessary to consider the relevance between the question and reviews, which can alleviate the noise from irrelevant review information.
(ii) Secondly, the \textit{textual similarity} between each pair of answers indicates their content consistency, which can be regarded as a notion of peer reviews among the entire answer set for verifying the reliability of each answer \cite{DBLP:conf/emnlp/TymoshenkoM18}.
Returning to the above example, since more answers agree that "the headphone is compatible with Android devices", the answer with the opposite opinion, such as A4, is thus less reliable than others. 
Similarly, measuring the similarity between reviews also helps cross verify the content consistency among them and hence captures the crowd's common opinions reflected in the review set for the given question. 
(iii) Thirdly, one may notice that such common opinions from the reviews often reveal authentic and general judgement from the whole community.
Thus, the relationship between an answer and reviews can be modeled by the \textit{textual entailment} relation \cite{DBLP:conf/emnlp/snli, DBLP:conf/acl/esim-17}, which examines whether the opinion holding by an answer is coherent with common opinions reflected in the reviews.
As shown in Table \ref{intro-example}, the review snippets reveal that "the concerned headphone can work with Android", indicating that the first three answers, i.e., A1, A2 and A3, are more consistent with the opinions from the whole community, which leads to a higher rank.
In summary, how to model and utilize the aforementioned multiple semantic relations among the question, answers, and relevant reviews for ranking the answers poses a main challenge.

In this paper, we propose an answer ranking model named MUSE for product-related questions, which comprehensively models \underline{\textbf{mu}}ltiple \underline{\textbf{se}}mantic relations among the question, answers, and relevant reviews. Concretely, we first conduct a word-to-word attention mechanism from the question to each individual answer during the answer encoding phase. 
Then the important information in the answer text and the relevance information with the question can be highlighted to obtain the textual features for each answer. 
Next, we construct a multi-semantic relation graph with the question, each answer and each review snippet as nodes. Then a customized graph convolutional neural network is designed for explicitly modeling the interrelationship between the nodes under multiple semantic relations. 
Precisely, for a specific answer, the textual features obtained in the earlier step are further refined by considering the semantic relevance with the question, the textual similarity with other answers and the textual entailment relation with relevant reviews. 
By modeling the relations between a given answer with other answers and relevant reviews, the coherence information between the concerned answer with the common opinion is accumulated to assist the ranking.
Finally, we adopt a joint loss function, combining both pointwise and listwise learning approaches, to consider a specific answer both locally and globally in the entire answer set.
To summarize, the contributions of this paper are as follows:
\begin{itemize}[leftmargin=*]
    \item We investigate the problem of ranking user-provided answers in E-commerce and propose a framework to jointly model multiple semantic relations including the semantic relevance between the question and answers, the textual similarity among answers, and the textual entailment between answers and reviews.
    
    \item We model both textual and interaction features of each answer to facilitate the ranking task. Importantly, a novel graph convolutional operation is designed to integrate the coherence information under different semantic relations. 
    
    \item Experimental results on real-world E-commerce data across three product domains show that our proposed MUSE model achieves superior performance on the concerned task.
\end{itemize}

\section{Related Work}
\textbf{Answer Ranking.} \; Answer selection has been extensively studied for solving the answer ranking problem in retrieval-based question answering systems as exemplified in the community question answering (CQA) \cite{DBLP:conf/semeval/NakovHMMMBV17} and factoid question answering~\cite{DBLP:conf/emnlp/WangSM07}. The research on answer ranking has evolved from early information retrieval (IR) research, which primarily focused on feature engineering with syntactic or lexical approaches~\cite{DBLP:conf/emnlp/WangSM07,DBLP:conf/emnlp/SeverynM13}. In recent years, deep learning based answer selection methods make several breakthroughs and become the mainstream approach to tackle the answer ranking task. Most of the existing neural models adopt Siamese architecture~\cite{DBLP:conf/sigir/SeverynM15,DBLP:conf/sigir/RaoHL17}, attentive architecture~\cite{DBLP:conf/acl/TanSXZ16} or compare-aggregate architecture~\cite{DBLP:conf/ijcai/WangHF17,DBLP:conf/iclr/Wang017,DBLP:conf/acl/re2-2019} for modeling the semantic relevance between the question and answer without heavy feature engineering. 
Additionally, some latest studies also learn to rank question-answer pairs from different perspectives such as utilizing external knowledge~\cite{DBLP:conf/coling/DengSYLDFL18}, extracting length-adaptive features \cite{DBLP:conf/sigir/ShaoCCR19}, modeling user expertise \cite{DBLP:conf/www/LyuOWSC19}, and measuring answer novelties \cite{DBLP:conf/sigir/OmariCRS16,DBLP:conf/www/HarelAAR19}.

There are also some works focusing on measuring the quality of answers or similar text content, then the predicted qualities can be used to rank them. \citet{DBLP:conf/sigir/answer-quality-2010} evaluate and predict answer qualities in the CQA platforms.
\citet{DBLP:conf/naacl/HalderKS19} propose a neural model to predict the quality of a response post to the original post, with the awareness of several previous posts in the discussion forum. 
In the E-commerce scenario, some studies \cite{DBLP:conf/www/FanFGSL19,DBLP:conf/www/ChenQYZHLB19} utilize product information such as product titles to predict the quality of a customer review. 
Recently, \citet{Zhang-WWW2020-rahp} utilize user reviews for identifying helpful answers in the product question answering forums while they neglect the interrelationships among answers.
In this work, we focus on the answer ranking problem in E-Commerce settings, where we wish to rank multiple user-provided answers for a product-related question with the help of relevant reviews.

\vskip 0.06in

\noindent \textbf{Product-related Question Answering.} \;
Recent years have witnessed several successful applications in product-related question answering (PQA) problem. 
Most of the existing studies exploit customer reviews as major~\cite{DBLP:conf/emnlp/YuZC12,DBLP:conf/wsdm/YuL18,DBLP:conf/aaai/0007GZZWZS19} or auxiliary resources~\cite{DBLP:conf/www/McAuleyY16,DBLP:conf/icdm/WanM16,DBLP:conf/wsdm/ChenLJZC19,DBLP:conf/wsdm/GaoRZZYY19} for providing responses to the given question. \citet{DBLP:conf/www/McAuleyY16} divide the question type into yes/no questions and open-ended questions and then tackle the yes/no type question as a classification task, aiming to predict the answer as "yes", "no" or "unsure" with the help of reviews. Following this direction, \citet{DBLP:conf/wsdm/YuL18} further consider the latent aspect information to improve the answer prediction performance. Some other works \cite{DBLP:conf/emnlp/YuZC12,DBLP:conf/aaai/0007GZZWZS19} adopt the retrieval-based methods to retrieve certain review snippets serving as the response. For example, \citet{DBLP:conf/aaai/0007GZZWZS19} propose a multi-task learning framework to identify reviews for a given question. Recently, some studies utilize the reviews to generate a sentence as the response based on the sequence-to-sequence architecture \cite{DBLP:conf/wsdm/ChenLJZC19,DBLP:conf/wsdm/GaoRZZYY19}. Although most of these models utilize the review information, it is often assumed that the user-written answers are unavailable, which is different from our concerned task to directly rank these answers for the given question.

\vskip 0.06in

\noindent \textbf{Graph Neural Networks.} \;
Graph Neural Networks (GNNs)~\cite{DBLP:conf/iclr/gcn-17} have been widely adopted to model graph structure data.
Some latest studies exploit GNN in the IR-related tasks, which constructs text-based graphs to model the structural relation beyond the context itself. 
\citet{DBLP:conf/cikm/LiQLYL19} propose a large-scale anti-spam method based on GCN for detecting the spam advertisements. \citet{DBLP:conf/sigir/SunTDDN19} propose a GCN encoder for keyphrase extraction that can effectively capture document-level word salience. \citet{DBLP:conf/ijcai/ChenGRHXGYZ19} develop heterogeneous graph attention networks (HGAT) for user profiling. In this work, we explore the utilization of relational GNN \cite{DBLP:conf/esws/relation-gcn-18} to model the interactions between information from different sources under different semantic relations in E-commerce settings.

\section{Model}
\begin{figure*}[h]
\setlength{\abovecaptionskip}{2pt}   
\setlength{\belowcaptionskip}{2pt}
  \centering
  \includegraphics[width=\textwidth]{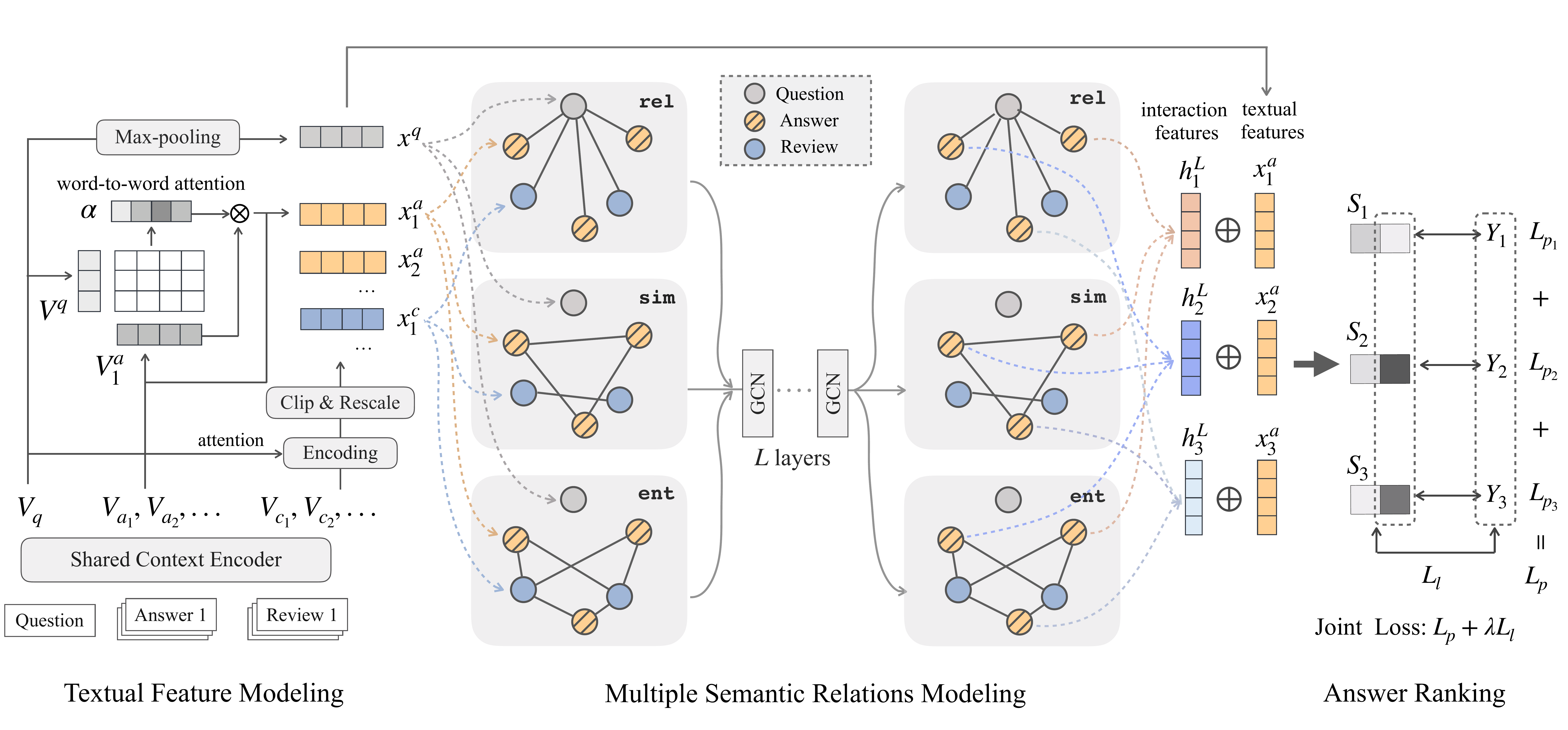}
  \caption{The architecture of the MUSE model composed of three main components, namely textual feature modeling, multiple semantic relations modeling, and answer ranking. Three types of semantic relations are specifically considered, including the semantic relevance ($\mathtt{rel}$) between the question with answers and with reviews, the textual similarity ($\mathtt{sim}$) among answers and among reviews, the textual entailment ($\mathtt{ent}$) between answers and reviews.} 
  \label{model}
  \vspace{-0.3cm}
\end{figure*}

In typical E-commerce settings, given a product-related question $q$ of a particular product, its answer set $\mathcal{A}=\{a_1, a_2, \dots\}$ contains $|\mathcal{A}|$ human-written answers. We can also obtain $|\mathcal{C}|$ relevant review snippets $\mathcal{C}=\{c_1, c_2, \dots \}$ to the question $q$. Our goal is to rank those answers in the answer set $\mathcal{A}$ with the review snippets $\mathcal{C}$ treated as auxiliary information that can be exploited to assist the ranking.

\subsection{Model Overview}
In this section, we introduce our proposed answer ranking model, MUSE, with modeling of multiple semantic relations among the question, answers, and relevant reviews. 
As shown in Figure \ref{model}, MUSE consists of three main components: textual feature modeling, multiple semantic relations modeling, and answer ranking.

We first employ a word-level attention to attend important and relevant information in the answers from the question during the textual feature modeling. Then a multi-semantic relation graph is constructed to model the multiple semantic relations among those texts from diverse information sources. Correspondingly, a customized graph convolutional network is developed to obtain the interaction-based features for each answer by aggregating the semantic relevance information between the question and answers, the textual similarity information among different answers, and the textual entailment information between the answer with each review snippet. Finally, after obtaining the textual features and interaction features, we design a joint loss function combining pointwise and listwise learning approaches to rank multiple answers.

\subsection{Textual Feature Modeling}

\subsubsection{\textbf{Context Modeling}}
Given a text sequence, which can be either the question sentence $q$, an answer sentence $a_i$, or a review snippet $c_i$, we first map each word in the sequence to a $d_e$-dimensional dense vector which can be initialized with pre-trained word vectors. We denote the embeddings for the word $w_i$ as $e_i \in \mathbb{R}^{d_e}$. To model the context interactions among words in the sequence, a bi-directional LSTM encoder is then employed to transform each word into a context-aware vector representation:
\begin{equation}
    v_i = \operatorname{Bi-LSTM}(e_i, v_{i-1}),
\end{equation}
where $v_i$ is the hidden state of the Bi-LSTM encoder at the $i$-th time step. We thus denote the representation of the question, the $i$-th answer $a_i$ and the $i$-th review snippet $c_i$ after such context-aware encoding as $V_q$, $V_{a_i}$ and $V_{c_i}$ respectively:
\begin{equation} \label{v-a}
    V_{*} = [v^*_1, v^*_2, ..., v^*_{|*|}]; \; \; * \in \{q, a_i, c_i\},
\end{equation}
\begin{equation}
    V_q \in \mathbb{R}^{|q| \times d_h}, V_{a_i} \in \mathbb{R}^{|a_i| \times d_h}, V_{c_i} \in \mathbb{R}^{|c_i| \times d_h},
\end{equation}
where $|q|$, $|a_i|$ and $|c_i|$ denote the sequence lengths of the corresponding text sequences, $d_h$ is the dimension of the hidden state of the Bi-LSTM encoder. To avoid notational clutter, we will omit the index of the answer and review snippet in this section since the same operations are conducted for each answer and review respectively. For example, we will use $V_a$ to represent the context-aware representation for one particular answer instead of $V_{a_i}$.

\subsubsection{\textbf{Question-attended Answer Encoding}}
To explicitly highlight the core semantic units in the answer sentence and their relevance with the question, we employ a word-to-word attention mechanism to attend the important information in the question to each word of the answer. Specifically, for the $i$-th word in $V_a$, we consider its similarity with every single word in the question:
\begin{gather}
    \alpha_i = \operatorname{tanh}(V_q \cdot v^a_i + b_a) \in \mathbb{R}^{|q|},\\
    \alpha_{i,j}' = \frac{\operatorname{exp}(\alpha_{i,j})}{\sum_k \operatorname{exp}(\alpha_{i,k})} ,\\
     o^a_i = \sum\nolimits_{j=1}^{|q|} \alpha_{i,j}' \cdot v^q_j \in \mathbb{R}^{d_h},
\end{gather}
where $\alpha_i$ denotes the similarity scores of the $i$-th word in the answer with every word in the question, $\alpha_{i, j}'$ is the normalized importance weight between the $i$-th word in the answer and the $j$-th word in the question. Then for each word in the answer, we obtain a question-attended representation $o^a_i$ as a weighted sum of the embeddings of question words. 
Next we concatenate the context-aware answer representation $v^{a}_i$ and the question-attended answer representation $o^{a}_i$ for the $i$-th word to obtain an enriched answer representation:
\begin{equation} \label{answer-distill}
    \hat{v^a_i} = \operatorname{tanh}(W_a \cdot [v^a_i; o^a_i] + b_{aa}),
\end{equation}
where $W_a$ and $b_{aa}$ are trainable parameters, [;] denotes the concatenation operation. A max-pooling operation is then employed to obtain the encoded vector representations $x_a$ and $x_q$ for the answer $a$ and the question $q$ respectively:
\begin{gather}
    x_a = \operatorname{Max-Pool}([\hat{v^a_1}, \hat{v^a_2}, \dots, \hat{v^a_{|a|}}]) \in \mathbb{R}^{d_h}, \\
    x_q = \operatorname{Max-Pool}([v^q_1, v^q_2, \dots, v^q_{|q|}]) \in \mathbb{R}^{d_h}.
\end{gather}
We denote the textual features obtained from the above operations for the $i$-th answer as $x^a_i$.

\subsubsection{\textbf{Clip-Rescale Attention for Review Encoding}}
For the relevant reviews, we also obtain encoded review representations as auxiliary information for assisting the ranking of answers. Although the review snippets in $\mathcal{C}$ are typically obtained with an initial retrieval process, there is still much noise contained in them, since these product reviews are originally written without explicitly responding to any question. To prevent the irrelevant information distracting the encoding, we employ a more aggresive clip-and-rescale attention mechanism inspired by \cite{dynamic-clip-17} to obtain the question-attentive representation for each review snippet:
\begin{gather}
    \beta = \operatorname{softmax}(V_{c} W_c x_q^T + b_c) \in \mathbb{R}^{|c|},\\
    \beta' = \operatorname{Rescale}(\beta \odot m) \in \mathbb{R}^{|c|},
\end{gather}
where $W_c$ and $b_c$ are trainable parameters, $V_c$ denotes the context-aware representation for a review snippet $c$, $\beta$ contains the original attention weights for each word in the review, $m = \{0, 1\} \in \mathbb{R}^{|c|}$ denotes a mask vector for $\beta$ where only the index whose corresponding weight score are among the top $k$ in $\beta$ will be $1$, and $0$ otherwise,  $\odot$ denotes the element-wise vector multiplication.  $\operatorname{Rescale()}$ refers to the vector rescale operation: $\operatorname{Rescale(v)} = v/\smallsum_{i} |v_i|$. Hence $\beta_j' \in \mathbb{R}$ refers to the importance score of the $j$-th word in the review snippet and is forced to be 0 for those unimportant words. Then we compute the review representation as the weighted sum of its context-aware representation: 
\begin{equation}
    x_{c} = \sum\nolimits_{i=1}^{|c|} \beta'_i v^c_i.
\end{equation}

The same operations introduced above are conducted for every review snippet. We thus denote the vector representation after such textual feature modeling for the $i$-th review as $x^c_i$. Following similar notation convention, we denote the question representation as $x^q$.

\subsection{Multiple Semantic Relations Modeling}
\subsubsection{\textbf{Multiple Semantic Relations}}
Effectively ranking the answers requires to exploit the rich semantic relations among the question, answers, and reviews. To this end, we identify three types of semantic relations among these diverse information sources which are useful for ranking the user-provided answers: 

\noindent \textbf{(1) Semantic Relevance} between the question and answer text is typically exploited in the general answer ranking task \cite{rao2019bridging}. After bringing in review information, measuring the semantic relevance between the question and review snippets is also useful for alleviating the noise from irrelevant review information.

\noindent \textbf{(2) Textual Similarity} between each pair of answers can effectively measure their content consistency \cite{DBLP:conf/emnlp/TymoshenkoM18} and hence help identify the core opinions in the entire answer set for the given question. Similarly, considering such relation among review snippets in the review set can reveal the common opinions reflected in the reviews.

\noindent \textbf{(3) Textual Entailment} relation between an answer and a review snippet indicates whether the answer is supported by that specific review, which is inspired by some attempts of utilizing textual entailment relation for general question answering problem \cite{DBLP:conf/acl/HarabagiuH06,DBLP:conf/naacl/TrivediKKSB19}. Concretely, we treat the review as external evidence to examine the opinion coherence of a given answer with the common opinions of the community.

To model these different semantic relations, especially capturing the coherence information of an answer with other answers and user reviews, it requires to aggregate the complex interactions among the existing information sources. Importantly, we can observe that these relations are closely connected and supposed to be modeled concurrently when ranking the answers. For example, each answer needs to be considered with different purposes when measuring its relation with the question, other answers, and review snippets. The coherence information from one relation can also affect its interaction with another information sources under different relations. Therefore, we propose a multi-semantic relation graph and utilize the graph convolutional networks (GNN) \cite{DBLP:conf/iclr/gcn-17}, which is shown to excel at aggregating the structural information from the neighborhoods, 
to capture the coherence information under different semantic relations. 

\subsubsection{\textbf{Graph Construction}}
Formally, we denote an undirected graph with multiple semantic relations as $\mathcal{G=(N, E, R)}$, with nodes $n_i \in \mathcal{N}$, labeled edges (i.e., semantic relations) between node $n_i$ and $n_j$ as $(n_i, r, n_j) \in \mathcal{E}$, where $r \in \mathcal{R}$ is the relation type between two nodes. Then to construct the graph, we treat the question $q$, each answer sentence $a_i \in \mathcal{A}$ and each review snippet $c_i \in \mathcal{C}$ as a node in $\mathcal{G}$. The total number of nodes is thus $1+|\mathcal{A}|+|\mathcal{C}|$.
We initialize each node with their corresponding textual features $x^*$ obtained from the textual feature modeling, which are encoded with their core semantic information.

To represent the multiple semantic relations, we make use of different adjacency matrices for the graph $\mathcal{G}$. Specifically, the relation type between two nodes $r \in \mathcal{R}=\{ \mathtt{rel,\, sim, \, ent}\}$, which represents the semantic relevance, textual similarity, and textual entailment relations respectively. Three adjacency matrices can thus be constructed for $\mathcal{G}$:
\begin{align}
 & A^{\mathtt{rel}}_{i,j} =
  \begin{cases}
    1       & \quad \text{if } n_i = q, n_j \in \{\mathcal{A}, \mathcal{C}\} \\
    0       & \quad \text{otherwise}
  \end{cases} \\
 & A^{\mathtt{sim}}_{i,j} =
  \begin{cases}
    1       & \quad \text{if } n_i, n_j \in \mathcal{A} \text{ or } n_i, n_j \in \mathcal{C} \\
    0       & \quad \text{otherwise}
  \end{cases} \\
 & A^{\mathtt{ent}}_{i,j} =
  \begin{cases}
    1       & \quad \text{if } n_i \in \mathcal{A}, n_j \in \mathcal{C} \\
    0       & \quad \text{otherwise}
  \end{cases}
\end{align}

\subsubsection{\textbf{Coherence Information Aggregation}}

Motivated by the Relational GCN \cite{DBLP:conf/esws/relation-gcn-18}, which shows good performance when considering the multiple relations between entities in a knowledge graph for the link prediction task, we develop a novel architecture for modeling the multiple semantic relations among the question, answers, and reviews for the concerned task. For a node $n_i$, the opinion coherence information is aggregated from its neighboring nodes:
\begin{equation} \label{gcn}
    h_{i}^{(l+1)} = \operatorname{ReLU}\left(\sum_{r \in \mathcal{R}} \sum_{j \in \mathcal{N}_{i}^{r}} \Lambda^r_{i, j} W_{r}^{(l)} h_{j}^{(l)}+W_{s}^{(l)} h_{i}^{(l)}\right),
\end{equation}
where $h_i^{(l)}$ is the hidden state of the node $n_i$ at the $l$-th layer of the network, $\mathcal{N}_{i}^{r}$ denotes the neighboring indices of the node $n_i$ under the relation $r$,  $W_{r}^{(l)}$ and $W_{s}^{(l)}$ are trainable parameters representing the transformation from neighboring nodes and from the node $n_i$ itself. $\Lambda^r_{i, j}$ is a normalization constant such as $\Lambda^r_{i, j}=1/|\mathcal{N}_{i}^{r}|$ in \cite{DBLP:conf/esws/relation-gcn-18}. To avoid the scale changing of the feature representation as commonly observed to be harmful for the performance, we apply a symmetric normalization transformation:
\begin{equation}
    \Lambda^r = D_r^{-1/2}A^{r}D_r^{-1/2}, \; r \in \{ \mathtt{rel, sim, ent}\},
\end{equation}
where $A^r$ is the adjacency matrix under the relation $r \in \mathcal{R}$, $D_r$ is the corresponding degree matrix of $A^r$. 

Unlike the basic GCNs using one convolutional filter matrix to model the feature transformation, the aggregation operation in Equation (\ref{gcn}) employs different weight matrices $W_r^{(l)}$ for different semantic relations in each layer, which can capture the coherence information explicitly under different relations. Besides, a self-connection weight matrix $W_s^{(l)}$ is utilized to control how much information in the node $n_i$ itself at each update should be kept. 

At each step, the representations of the answer nodes are enriched by their neighbourhoods. 
Specifically, the first layer takes the textual feature vectors obtained from Section 3.2 as the input for each node, i.e. for the question node: $h^{(0)}=x^q$, for the node of the $i$-th answer: $h^{(0)}=x^a_i$ and for the node of the $i$-th review snippet: $h^{(0)}=x^c_i$. 
Then the transformation in Equation (\ref{gcn}) can be stacked up to $L$ layers to include the dependencies across multiple relational steps. 
We take the output of each answer node at the last layer as their interaction features and denote $h^L_i$ as the feature representation for the $i$-th answer.

\subsection{Answer Ranking}
For each answer $a_i$, after obtaining the textual feature $x^a_i$ and the interaction features $h^L_i$, they are then concatenated and fed to a MLP with one hidden layer to get the final prediction scores:
\begin{equation} \label{obtain-si}
    S_i = \operatorname{MLP}([x^a_i; h_i^L]) \in \mathbb{R}^2,
\end{equation}
where $S_i$ is the prediction vector for the $i$-th answer. We denote the concatenation of the prediction vectors of the entire answer set as $S \in \mathbb{R}^{|\mathcal{A}| \times 2}$. Finally, we employ a joint loss function, combining the pointwise and listwise learning approaches, to conduct training for learning to rank the answers.

\subsubsection{\textbf{Pointwise Loss Function}}
One of the most commonly used training strategies for answer ranking problem is the pointwise learning approach. Specifically, for each answer $a_i$, a softmax function is applied to its prediction vector $S_i$ to obtain the predicted distribution $\hat{S_i} = \operatorname{Softmax}(S_i) \in \mathbb{R}^2$. Then each answer is considered separately and the cross-entropy loss is computed:
\begin{equation} \label{pointwise}
    \mathcal{L}_p = \frac{1}{|\mathcal{A}|} \sum\nolimits_{i=1}^{|\mathcal{A}|} (- Y_i \; log \; \hat{S}_i),
\end{equation}
where $Y_i \in \mathbb{R}^2$ denotes the one-hot encoding of the label for the $i$-th answer. Then the total loss $\mathcal{L}_p$ for each question is the average of the cross-entropy loss of all its answers.

\subsubsection{\textbf{Listwise Loss Function}}
Another choice of learning approach is the listwise method considering all candidate answers to the given question at the same time. Given the prediction matrix $S$, we first normalize the prediction scores among all answers:
\begin{equation}
    \hat{y} = \frac{[S_{1,1}, S_{2,1}, ..., S_{|\mathcal{A}|,1}]}{\|[S_{1,1}, S_{2,1}, ..., S_{|\mathcal{A}|,1}]\|_p} \in \mathbb{R}^{n_a},
\end{equation}
where $S_{i,1}$ is the unnormalized prediction score of answer $a_i$ being a positive answer, $p$ is set to $1$ in the experiments to compute the vector norm. Similarly, we also normalize the whole label list $y \in \mathbb{R}^{|\mathcal{A}|}$ of all answers with $y' = y / \| y \|_p$ where $y_i =\{0, 1\}$ is the raw label for the $i$-th answer. Then we can compute the listwise loss for a given question with Kullback-Leibler divergence: 
\begin{equation} \label{listwise}
    \mathcal{L}_l = \frac{1}{|\mathcal{A}|} KL\_ Div(\hat{y} || y' ) = \frac{1}{|\mathcal{A}|} \sum\nolimits_{i=1}^{|\mathcal{A}|} \hat{y}_i \; log (\frac{\hat{y}_i}{y'_i}).
\end{equation}

\subsubsection{\textbf{Joint Loss Function}}

The pointwise and listwise learning approaches consider a specific answer locally and globally in the entire answer set respectively. In the pointwise learning approach, we focus on each individual answer locally, and the goal is to accurately predict their corresponding labels. In the listwise method, we examine the entire answer list globally, attempting to differentiate the good and bad answers, and making the former rank higher. In order to combine the strengths of both of them, we propose to employ a joint loss function in this work.

Specially, we combine these two types of loss functions to a joint loss $\mathcal{L}$ to train our proposed MUSE model. The above introduced two loss functions in Equation (\ref{pointwise}) and Equation (\ref{listwise}) are for one single question (i.e. one data instance) in the dataset. Then for the whole dataset with in total $|\mathcal{Q}|$ questions, the joint loss function is:
\begin{equation} \label{joint-loss-eq}
    \mathcal{L} = \frac{1}{|\mathcal{Q}|} \sum\nolimits_{i=1}^{|\mathcal{Q}|} ( \mathcal{L}_{p_i} + \lambda \mathcal{L}_{l_i} ) + \eta \; \| \Theta \|_2,
\end{equation}
where $\mathcal{L}_{p_i}$ and $\mathcal{L}_{l_i}$ are the pointwise and listwise loss functions for the $i$-th question respectively. $\lambda$ is a hyper-parameter for balancing between these two loss functions, $\eta$ is the L2 regularizer weight, $\Theta$ is the set of all trainable parameters in the model.

\section{Experiment Setup}
\subsection{Datasets}

We evaluate our proposed model on Amazon QA dataset \cite{DBLP:conf/icdm/WanM16}, and utilize three product categories with the largest number of question-answer pairs for evaluation, including \textit{Electronics, Home and Kitchen, Sports and Outdoors}.
The dataset contains questions accompanied with their multiple user-written answers. The product ID of each question is then utilized to align with the Amazon review dataset \cite{DBLP:conf/www/amazon-review-16,DBLP:conf/emnlp/NiLM19} for obtaining the corresponding reviews for the question. Since an entire raw review can be lengthy and talks about multiple aspects of the concerned product, each review text is chunked into snippets at the sentence level. Then for each question, we adopt BM25 to rank all the review snippets and collect the top 5 relevant snippets for each question in our experiments.

Similar to previous works \cite{DBLP:conf/naacl/HalderKS19,Zhang-WWW2020-rahp}, we treat the user votes from the community as a proxy of the gold label for the quality of an answer. Thus, the answer whose number of positive votes is greater than the number of negative votes is treated as a high-quality (positive) answer, otherwise it is treated as a negative one. We split 10\% of the dataset for each product category for testing and the rest is used for training and validation. The statistics are summarized in Table \ref{dataset}, including the number of products (\# Product), questions (\# Q), answers (\# A), and positive answers (\# Pos A).

\begin{table}
\setlength{\abovecaptionskip}{2pt}%
\setlength{\belowcaptionskip}{2pt}
\fontsize{8.5}{10.5}\selectfont
  \caption{Statistics of data splits of three product categories.}
  \label{dataset}
  \begin{tabular}{c|c|cccc}
    \toprule
    \multicolumn{2}{c}{Category} & \# Product & \# Q & \# A & \# Pos A  \\
    \midrule
    \multirow{2}{*}{Electronics} & Train+Val & 11,172 & 15,547 & 80,115 & 28,919 \\
    & Test & 1,657 & 1,727 & 8,823 & 3,184 \\
    \midrule
    \multirow{2}{*}{Home} & Train+Val & 8,590 & 12,731 & 66,956 & 24,838 \\
    & Test & 1,349 & 1,414 & 7,461 & 2,801 \\
    \midrule
    \multirow{2}{*}{Sports} & Train+Val & 4,949 & 6,952 & 35,858 & 13,230 \\
    & Test & 746 & 772 & 4,065 & 1,511 \\
    \bottomrule
  \end{tabular}
\vspace{-0.3cm}
\end{table}

\begin{table*}
\centering
\setlength{\abovecaptionskip}{2pt}   
\setlength{\belowcaptionskip}{2pt}
  \caption{Answer ranking results of MUSE and baseline models. 
  $\dagger$ denotes that MUSE-Joint-Loss model achieves better performance than the strong baseline PHP with statistical significance test for $p < 0.05$.}
  \label{results}
  \begin{tabular}{c|cccc|cccc|cccc}
    \toprule
      & \multicolumn{4}{c}{Electronics} 
      & \multicolumn{4}{c}{Home \& Kitchen}
      & \multicolumn{4}{c}{Sports \& Outdoors} \\
    
    \cmidrule(lr){2-5} \cmidrule(lr){6-9} \cmidrule(lr){10-13} 
    
    & MAP & MRR & P@1 & P@3 & MAP & MRR & P@1 & P@3 & MAP & MRR & P@1 & P@3 \\
    \midrule
    BM25 \cite{DBLP:journals/ftir/bm25} & 0.571 & 0.576 & 0.380 & 0.383 & 0.585 & 0.592 & 0.406 & 0.397  & 0.586 & 0.589 & 0.384 & 0.398  \\
    CNN \cite{DBLP:conf/sigir/SeverynM15}  & 0.633 & 0.668 & 0.460 & 0.411 & 0.638 & 0.675 & 0.474 & 0.407 & 0.628 & 0.664 & 0.452 & 0.407 \\
    aNMM \cite{DBLP:conf/cikm/anmm}  & 0.619 & 0.651 & 0.445 & 0.386 & 0.633 & 0.670 & 0.465 & 0.413 & 0.624 & 0.659 & 0.448 & 0.403 \\
    Att-BiLSTM \cite{DBLP:conf/acl/TanSXZ16} & 0.642 & 0.671 & 0.464 & 0.408 & 0.639 & 0.673 & 0.471 & 0.416 & 0.633 & 0.665 & 0.464 & 0.408 \\
    BiMPM \cite{DBLP:conf/ijcai/WangHF17}  & 0.647 & 0.678 & 0.480 & 0.405 & 0.656 & 0.688 & 0.491 & 0.425 & 0.636 & 0.680 & 0.482 & 0.409  \\
    HCAN \cite{rao2019bridging}  & 0.643 & 0.676 & 0.472 & 0.412 & 0.659 & 0.686 & 0.492 & 0.429 & 0.632 & 0.666 & 0.459 & 0.404  \\
    PRHNet \cite{DBLP:conf/www/FanFGSL19} & 0.646 & 0.677 & 0.478 & 0.406 & 0.649 & 0.683 & 0.483 & 0.421 & 0.634 & 0.669 & 0.469 & 0.405 \\
    PHP \cite{DBLP:conf/naacl/HalderKS19}  & 0.652 & 0.679 & 0.475 & 0.414 & 0.648 & 0.681 & 0.484 & 0.421 & 0.638 & 0.667 & 0.463 & 0.409 \\
    \midrule
    MUSE-Pointwise & 0.663 & 0.693 & 0.504 & 0.425  & 0.679 & 0.710 & 0.521 & 0.450  & 0.649 & 0.684 & 0.482 & 0.431  \\
    MUSE-Listwise & 0.678 & \textbf{0.715} & \textbf{0.539} & 0.417 & 0.675 & 0.712 & \textbf{0.527} & 0.443 & 0.657 & 0.688 & 0.491 & 0.435 \\
    MUSE-Joint-Loss  & \textbf{0.695}$\dagger$ & 0.711$\dagger$ & 0.511$\dagger$ & \textbf{0.450}$\dagger$ & \textbf{0.693}$\dagger$ & \textbf{0.714}$\dagger$ & 0.518$\dagger$ & \textbf{0.466}$\dagger$ & \textbf{0.661}$\dagger$ & \textbf{0.694}$\dagger$ & \textbf{0.498}$\dagger$ & \textbf{0.437}$\dagger$ \\
    \bottomrule
  \end{tabular}
\end{table*}

\subsection{Baselines and Evaluation Metrics}
We compare our proposed MUSE model with some traditional and state-of-the-art methods.
To conduct a more comprehensive comparison, we slightly modify some models to take the advantage of utilizing relevant reviews as one of their inputs.
\begin{itemize}
    \item \textbf{BM25} \cite{DBLP:journals/ftir/bm25}: It is a widely-used retrieval model for ranking candidate answers given a question.
    \item \textbf{CNN} \cite{DBLP:conf/sigir/SeverynM15}: It employs a CNN-based Siamese network to encode QA pairs for ranking the answers.
    \item \textbf{Attentive-BiLSTM} \cite{DBLP:conf/acl/TanSXZ16}: It utilizes a bidirectional LSTM as well as an attention mechanism to measure the relevance between the question and answer text.
    \item \textbf{aNMM} \cite{DBLP:conf/cikm/anmm}: It is an \textbf{a}ttention based \textbf{N}eural \textbf{M}atching \textbf{M}odel,
    which employs a value-shared weights scheme and a gated attention network to improve the ranking performance.
    \item \textbf{BiMPM} \cite{DBLP:conf/ijcai/WangHF17}: \textbf{Bi}lateral \textbf{M}ulti-\textbf{P}erspective \textbf{M}atching is one of the state-of-the-art models in many retrieval based QA tasks. It matches QA sentence pair from multiple perspectives.
    \item \textbf{HCAN} \cite{rao2019bridging}: \textbf{H}ybrid \textbf{C}o-\textbf{A}ttention \textbf{N}etwork is one recent model for modeling short text relations. It combines the semantic matching and relevance matching components to complement each other for better performance.
    \item \textbf{PRHNet} \cite{DBLP:conf/www/FanFGSL19}: It is one of the state-of-the-art models for predicting the quality of product reviews. We concatenate the QA pair and treat it as a single review and utilize relevant reviews as the "product information" in the original model.
    \item \textbf{PHP} \cite{DBLP:conf/naacl/HalderKS19}: \textbf{P}ost \textbf{H}elpfulness \textbf{P}rediction is one recent model for predicting the quality of a replying post to an initial post in the online discussion forum. We treat the question and answer as the original and replying post respectively, and treat reviews as the previous posts used in the model.
\end{itemize}

For our proposed MUSE model, we use different suffixes to denote the variants of it trained with different learning approaches, where "-Pointwise", "-Listwise" and "-Joint-Loss" refer to training with the pointwise, listwise, and joint loss function, respectively. 

For evaluation metrics, Mean Average Precision (MAP), Mean Reciprocal Rank (MRR), as well as Precision@N (P@N) are adopted to measure the model performance on ranking user answers. 
We take $N=1$ and $N=3$ in the experiments for P@N metric, which correspond to two common use cases in E-commerce: 
P@1 measures the performance when only a single answer is paired with each question in the product main 
page\footnote{e.g.,   \url{https://www.amazon.com/dp/B07DLPWYB7?th=1\#Ask}} and 
P@3 measures how well the top three answers for each question in the detailed question
page containing all user-provided answers\footnote{e.g.,  \url{https://www.amazon.com/ask/questions/Tx12DHUXVP6P535/ref=ask_ql_ql_al_hza}}.

\subsection{Experiment Configurations}
For the network configurations, we tuned hyper-parameters with the validation set. 
Specifically, the hidden dimension of the Bi-LSTM context encoder is set to 100, the hidden size of the weight matrix $W_a$ in Equation (\ref{answer-distill}) is 200, $k$ is set to 8 for alleviating the noise in reviews. 
In the semantic relation modeling, we stack two GCN layers to obtain the interaction features for the answers, i.e. $L=2$, where the hidden dimension of each layer is set to 150 and 100 respectively. The hyper-parameter $\lambda$ used to balance between two loss functions in Equation (\ref{joint-loss-eq}) is set to 2 and $\eta$ is set to 0.001.

For the training process, we initialize the word embedding layers of all neural models with the pre-trained 300D GloVE word embeddings\footnote{\url{http://nlp.stanford.edu/data/glove.6B.zip}}.
We adopt the Adam optimizer \cite{kingma2014adam} to train all learnable parameters and the batch size is set to 50.
All the network weights $W_*$ are initialized randomly from Xavier uniform distribution \cite{DBLP:journals/jmlr/xavier}.

\section{Results and Analysis}

\begin{figure*}
     \centering
     \begin{subfigure}[b]{0.5\textwidth}
         \centering
         \includegraphics[width=0.9\linewidth]{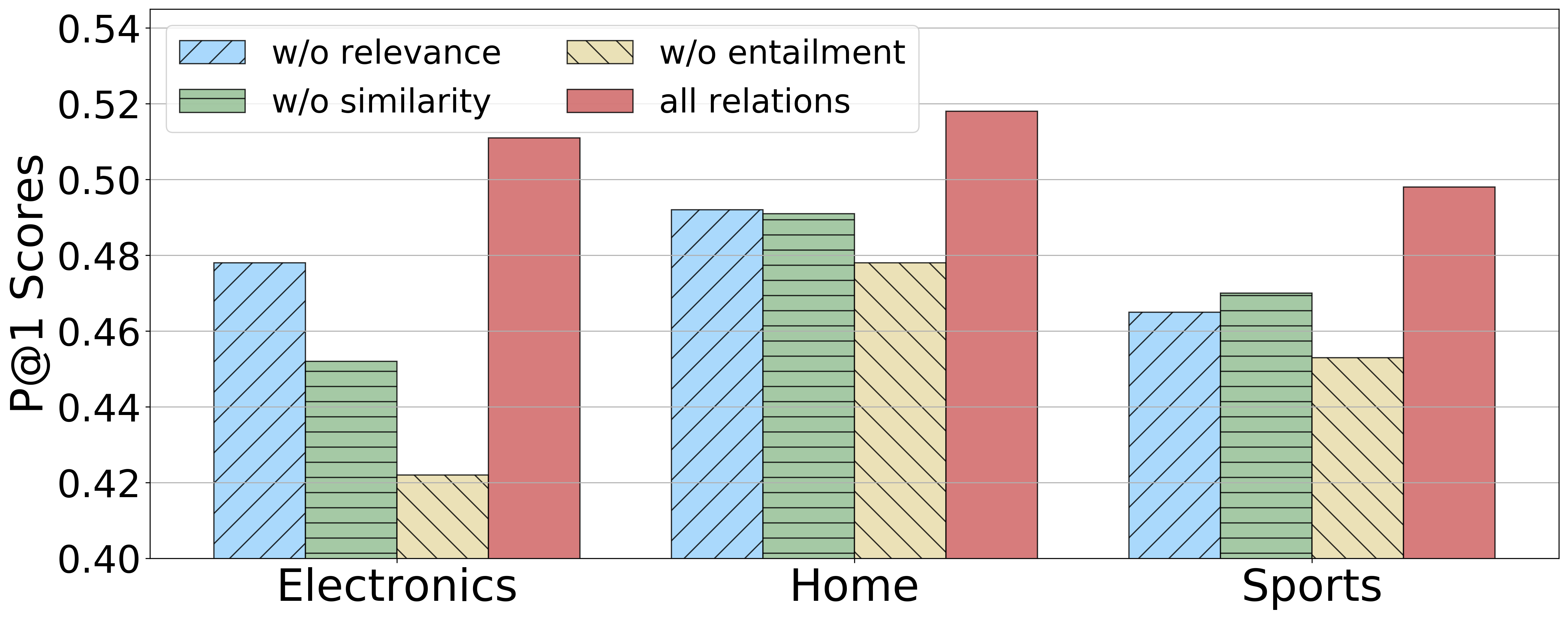}
         \caption{P@1 scores for three datasets}
         \label{p-at-1}
     \end{subfigure}%
     \begin{subfigure}[b]{0.5\textwidth}
         \centering
         \includegraphics[width=0.9\linewidth]{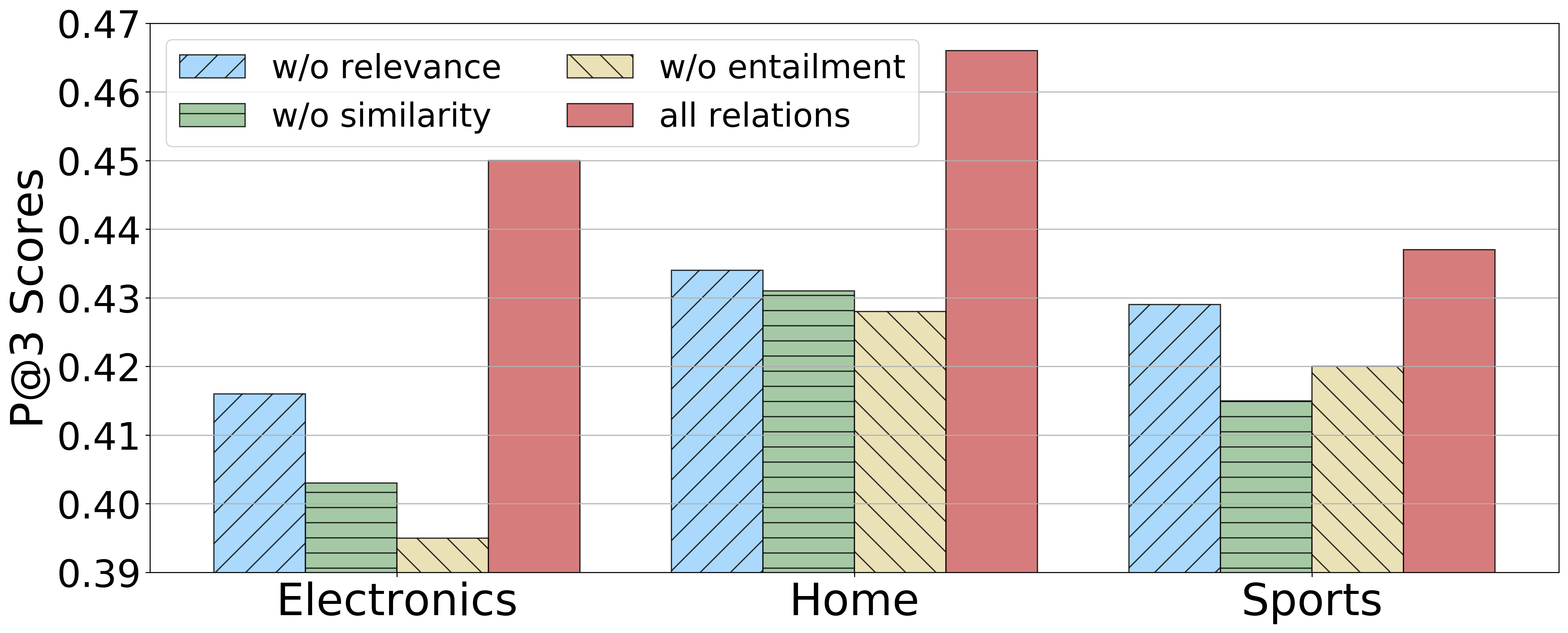}
         \caption{P@3 scores for three datasets}
         \label{p-at-3}
     \end{subfigure}
     \caption{Effect of different semantic relations in terms of P@1 and P@3 scores among three product categories}
     \label{graph-relations}
\end{figure*}

\subsection{Answer Ranking Performance}
The answer ranking results among three product categories in terms of MAP, MRR, P@1, and P@3 scores are summarized in Table \ref{results}. It shows that MUSE outperforms all baseline models on each dataset. 
Also, we conduct a statistical significance test comparing MUSE with PHP. The results indicates that MUSE achieves better performance than PHP with statistical significance test at $p<0.05$.

There are several notable observations from the results: 
(1) Compared to the basic BM25 model, we can see that deep learning models generally provide strong baselines for the concerned ranking task. In particular, BiMPM and HCAN models outperform other answer selection models by taking into account deeper and boarder semantic relevance information.
(2) Models with consideration of relevant review information, e.g. PHP model and our proposed MUSE model, can generally achieve better performance. 
Such results demonstrate that merely considering the semantic relevance between the question and answer text is not sufficient for ranking the user-provided answers in E-commerce settings.
(3) Our proposed MUSE model consistently and substantially outperforms all the baselines across three categories. 
This result shows that carefully modeling the rich semantic relations among the available information sources, i.e. the question, multiple answers, and relevant reviews, is necessary for effectively ranking the answers. Importantly, MUSE utilizes the multi-semantic relation graph to model the coherence information between each specific answer with the common opinions reflected in the entire answer set and reviews, which leads to its superior performance when ranking user answers in E-commerce scenario.

Comparing the performance between different variants of MUSE model, it can be observed that training with the joint loss function (i.e. MUSE-Joint-Loss) generally achieves better performance than learning with the pointwise approach (i.e. MUSE-Pointwise) or the listwise approach (i.e. MUSE-Listwise).
MUSE-Joint-Loss model combines the advantages of two learning approaches and considers an answer both from the perspective of its own label and from the perspective of the labels of the entire answer list, thus it achieves better results among the majority of cases.
Notably, MUSE-Listwise largely outperforms existing models regarding the MRR and P@1 scores. 
For example, it obtains about 6\% absolute improvement of P@1 score on the \textit{Electronics} category compared with the best performance given by the baseline model. 
The reason is that by normalizing the prediction list of all answers and minimizing its difference with the label list, we push the positive answers to have relatively higher scores. Thus the model can find one positive answer more easily and rank it as the top answer, which leads to better MRR and P@1 metrics.

\subsection{Ablation Study}

\begin{table}
\centering
\setlength{\abovecaptionskip}{2pt}   
\setlength{\belowcaptionskip}{3pt}
  \caption{Ablation study on the \textit{Electronics} category}
  \label{ablation}
  \begin{tabular}{l|cccc}
    \toprule
    Model Variants & MAP & MRR & P@1 & P@3 \\
    \midrule
    MUSE-Joint-Loss & \textbf{0.695} & \textbf{0.711} & \textbf{0.511} & \textbf{0.450} \\
    - w/o textual feature & 0.649 & 0.681 & 0.478 & 0.409 \\
    - w/o interactive feature & 0.646 & 0.673 & 0.468 & 0.411 \\
    - w/o Q-to-A attention & 0.656 & 0.688 & 0.490 & 0.414 \\
    - w/o Q-to-C attention & 0.663 & 0.692 & 0.497 & 0.419 \\
    \bottomrule
  \end{tabular}
\vspace{-0.3cm}
\end{table}

\subsubsection{\textbf{Impact of Main Components of MUSE}} 
We perform ablation studies by leaving out some important components in the proposed MUSE model to investigate their effectiveness. 
The results on the largest \textit{Electronics} dataset are presented in Table \ref{ablation}. We first create two variant models by discarding the textual feature $x_i$ (denoted as w/o textual feature) and the interaction feature $h_i$ (denoted as w/o interaction feature) for a specific answer when obtaining the prediction score $S_i$ in Equation (\ref{obtain-si}) respectively. From the results, we can find that both of them play an important role in contributing useful answer representations for the ranking. 
Specifically, the model without interaction feature suffers a slightly larger performance decrease, which is likely due to the fact that the textual features $x_i$ is used as the initialization to compute $h_i$ in the multiple semantic relation modeling so that some textual information encoded in $x_i$ can be preserved in the ranking process.

In addition, to testify the usefulness of the question attention operation during the textual feature modeling of the answers and reviews, we create two variant models by directly conducting a max-pooling operation on $V_a$ and $V_c$ in Equation (\ref{v-a}) to get $x_a$ and $x_c$ respectively, instead of employing attention mechanism (denoted as "w/o Q-to-A attention" and "w/o Q-to-C attention" in Table \ref{ablation} respectively). It can be observed that both lead to a performance decrease, indicating that utilizing the question to attend the important information during the encoding phase of the answer and review sentences is useful for capturing relevant and important information for the subsequent learning process.

Especially, we can notice that leaving out the question attention of the answers results in a larger performance decrease. This result shows that the core semantic information in answers is still essential for modeling the textual representations for answer ranking.

\subsubsection{\textbf{Impact of Different Semantic Relations.}} To better rank the multiple answers for a given question, we model the multiple semantic relations among the question, answers, and relevant reviews in this paper. 
Thus, we examine the effect of each semantic relation during the graph construction phase in this section by removing one relation at each time. 
We present P@1 and P@3 scores among three product categories in Figure \ref{graph-relations}, 
where "w/o relevance", "w/o similarity" and "w/o entailment" denote the MUSE-Joint-Loss model without the semantic relevance, textual similarity, and textual entailment relation when constructing the multi-semantic graph in Section 3.3.2 respectively. Also "all relation" refers to the performance of the model with all three relations. 
We can see that each of the semantic relations contributes to the final ranking performance and discarding any of them leads to performance degradation. 
This result illustrates the importance of explicitly modeling the complex relations among the multiple information sources in E-commerce scenario. 
In addition, it can be noticed that the entailment relation attaches more importance than the other two relations, 
which validates the necessity to utilize relevant reviews as external sources for modeling the opinion coherence between a concerned answer with the common opinion.
Moreover, discarding the semantic relevance relation leads to the least performance decrease since the reviews obtained by an initial retrieval process are somehow already related to the question.

\subsection{Analysis of MUSE model}
\begin{figure*}
     \setlength{\abovecaptionskip}{2pt}   
     \setlength{\belowcaptionskip}{2pt}
     \centering
     \begin{subfigure}[b]{0.7\textwidth}
         \centering
         \includegraphics[width=0.95\linewidth]{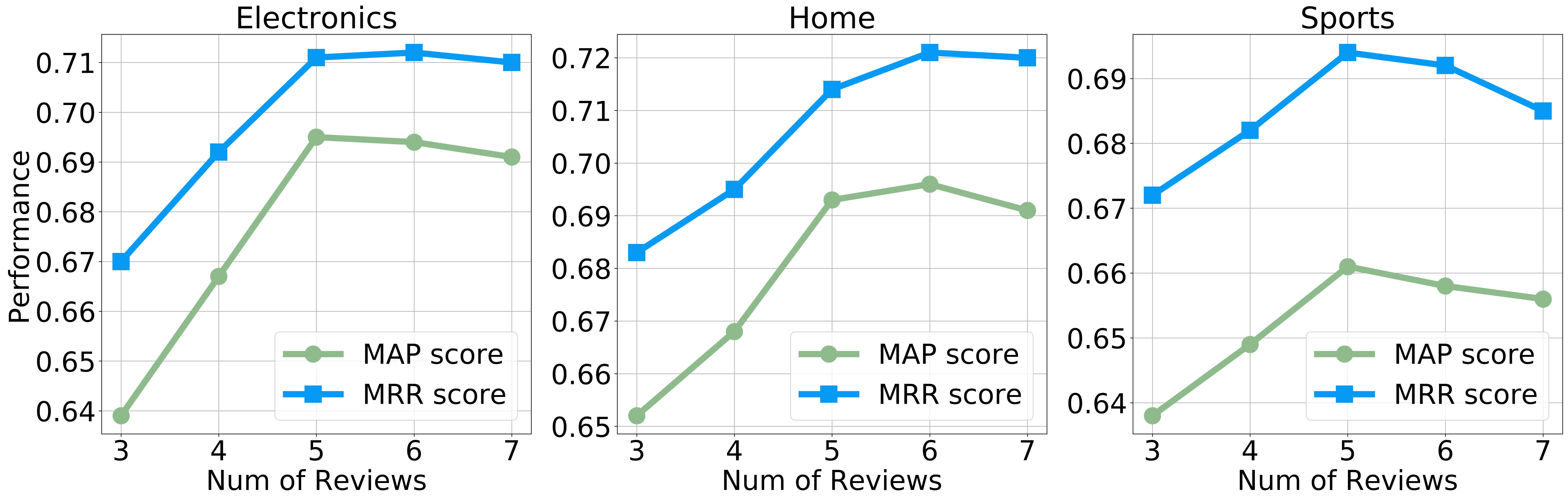}
         \caption{}
         \label{num-reviews}
     \end{subfigure}%
     \begin{subfigure}[b]{0.3\textwidth}
         \centering
         \includegraphics[width=0.95\linewidth]{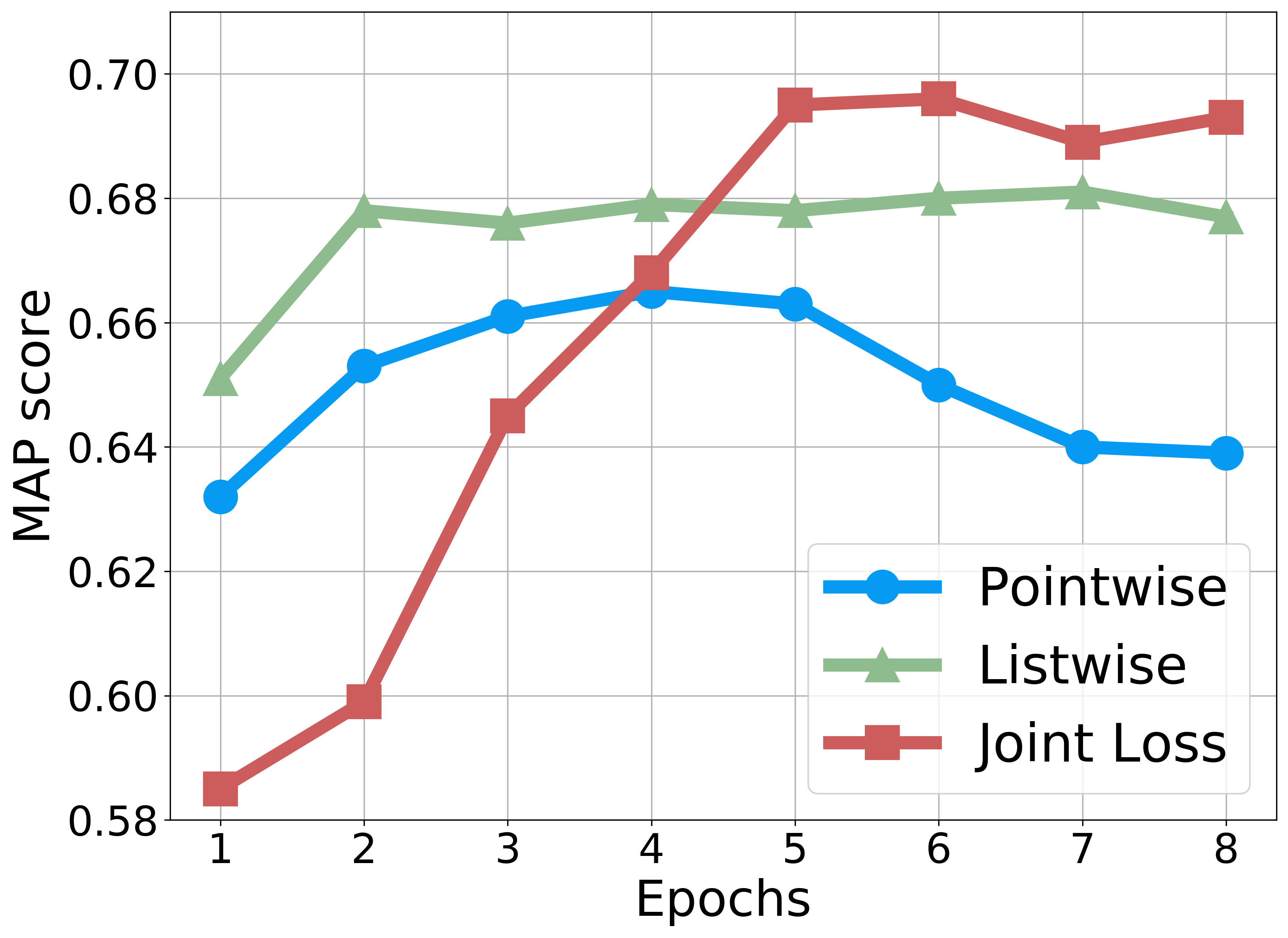}
         \caption{}
         \label{map-plw}
     \end{subfigure}
     \caption{Analysis of MUSE model. Figure (a) shows the performance with respect to the number of reviews utilized by MUSE on three datasets. Figure (b) presents the MAP score on the test set at each epoch of MUSE with different learning approaches.}
     \label{graph-relations}
\end{figure*}

\subsubsection{\textbf{Number of Reviews}} 

The relevant reviews are utilized as important external information in our concerned task. The proposed MUSE model aggregates common opinions with the review-review similarity relation and models the opinion coherence of an answer with the review-answer entailment relation. In this section, we vary the number of review snippets used in the model, i.e. the value of $|\mathcal{C}|$, to investigate its effect on the model performance. The MAP and MRR scores with different number of reviews used in the model on three datasets are presented in Figure \ref{num-reviews}. 

We can see that, as expected, the performance of the model is getting better when more review information is utilized at the beginning. However, both MAP and MRR scores become generally unchanged (e.g. on the \textit{Electronics} and \textit{Home} dataset) or even slightly decrease (e.g. on the \textit{Sports} dataset) when we further increase the number of reviews. 
On one hand, more reviews can provide more comprehensive common opinions from the community to help rank the candidate answers, leading to the performance improvement at the beginning. On the other hand, increasing the number of reviews used in the model also introduces more parameters and leads to the overfitting issue for those smaller datasets such as \textit{Sports}.

\subsubsection{\textbf{Different Learning Approaches.}}

We compute the MAP scores at each epoch on the test set of the proposed MUSE model trained with different loss functions. The results on the largest \textit{Electronics} dataset are shown in Figure \ref{map-plw}. 
It can be observed that MUSE trained with listwise loss function converges in fewer epochs than the other two approaches and is less affected by the overfitting issue, which is consistent with some observations from previous studies \cite{dynamic-clip-17, DBLP:journals/ftir/Liu09}. On the contrary, the model trained with pointwise learning approach is more likely to overfit after a few epochs. Such a phenomenon is likely due to the imbalance proportion between the high quality and low quality answers. Thus the model which is trained to only recognize the label of each answer individually will tend to predict an answer as a negative one so as to minimize the overall cross-entropy loss. For the model trained with the joint loss function, it is robust to the overfitting problem but converges at a relatively slow pace compared with the listwise learning approach.

\subsection{Case Study}
\begin{table}
\fontsize{8}{9}\selectfont
\setlength{\abovecaptionskip}{2pt}   
\setlength{\belowcaptionskip}{2pt}
  \centering
  \caption{A sample case of multiple answers ranked by their original community votes, as well as their predicted ranks by MUSE and two baseline models.}
  \label{case}
  \begin{tabular}{p{5cm}ccc}
    \toprule
    \multicolumn{4}{l}{\textbf{Question}: \; Is it automatic shut off?} \\ \midrule
    \multicolumn{4}{l}{\textbf{Relevant Review Snippets $\mathcal{C}$:}} \\
    \multicolumn{4}{l}{$c_1$: A buzzer sounds to let you know the eggs are done} \\
    \multicolumn{4}{l}{$c_2$: When the alarm sounds you need to turn it off and open it...} \\
    \multicolumn{4}{l}{$c_3$: I would have liked the cooker to turn off automatically but instead a} \\
    \multicolumn{4}{l}{\;\;\;\; bell rings until you turn if off.} \\
    \multicolumn{4}{l}{$c_4$: Also, by the time the timer goes off, the hot pan has a burning smell.} \\
    \multicolumn{4}{l}{$c_5$: and it turns off itself after the bell rings.} \\
    \midrule 
    \hfil \textbf{Answers} & \textbf{HCAN} & \textbf{PHP} & \textbf{MUSE} \\
    \midrule
    $a_1$: No, it beeps until you turn it off.  
    & R3 & R1 & R1   \\
     \midrule
    $a_2$: No it's not but it beeps very loud.  
    & R4 & R4 & R2   \\ 
    \midrule
    $a_3$: Yes, and it works very well.......we just had poached eggs yesterday. recommend this product :)  
    & \multirow{3}{*}{R1} & \multirow{3}{*}{R3} & \multirow{3}{*}{R3}  \\
    \midrule
    $a_4$: Yes there's an automatic shut-off when the cooking cycle is finished. 
    & \multirow{2}{*}{R2} & \multirow{2}{*}{R2} & \multirow{2}{*}{R4}  \\
  \bottomrule
\end{tabular}
\end{table}

To gain some insights into the proposed MUSE model, we present a sample case in Table \ref{case}, which includes a question of an \textit{egg cooker} product, its multiple user-provided answers, as well as the relevant review snippets. 
The answers $a_1,..a_4$ are ranked by their original community votes. We also present the ranks given by MUSE and two strong baseline models, namely HCAN and PHP for each answer, where "Rx" denotes that the answer is ranked at the $x$-th position by the corresponding model.
From the results, we can see that HCAN performs poorly on this case since it only considers the semantic relevance between the answer with the question text, while all the associated answers are quite topically relevant to the given question.
Besides, the PHP model, which incorporates review information into the modeling, also fails to ranks all answers correctly, indicating that the semantic relations need to be appropriately exploited. 
The proposed MUSE model utilizes the answer-answer similarity and review-review similarity relations to capture the common opinion that "the concerned egg cooker cannot turn off automatically". The answer-review entailment relation can then help examine the opinion coherence between each specific answer with the common opinion and hence help the ranking process. Therefore, it successfully ranks all answers in this case. This real-world example indicates the importance of taking review information into consideration. More importantly, carefully modeling the complex semantic relations between the question, answers, and reviews is essential for tackling this task in E-commerce settings.

\section{Conclusions}
We investigate the answer ranking problem for product-related questions in this paper. To tackle the ranking task in E-commerce settings, we propose a framework named MUSE to jointly model the multiple semantic relations among the question, answers, and relevant reviews. MUSE employs a novel graph convolutional operation customized to integrate the coherence information under different semantic relations to facilitate the ranking task. Extensive experiments on real-world E-commerce datasets show that our proposed model achieves superior performance compared with some strong baseline models.

\clearpage

\bibliographystyle{ACM-Reference-Format}
\balance
\bibliography{sample-base}


\begin{thebibliography}{54}


\ifx \showCODEN    \undefined \def \showCODEN     #1{\unskip}     \fi
\ifx \showDOI      \undefined \def \showDOI       #1{#1}\fi
\ifx \showISBNx    \undefined \def \showISBNx     #1{\unskip}     \fi
\ifx \showISBNxiii \undefined \def \showISBNxiii  #1{\unskip}     \fi
\ifx \showISSN     \undefined \def \showISSN      #1{\unskip}     \fi
\ifx \showLCCN     \undefined \def \showLCCN      #1{\unskip}     \fi
\ifx \shownote     \undefined \def \shownote      #1{#1}          \fi
\ifx \showarticletitle \undefined \def \showarticletitle #1{#1}   \fi
\ifx \showURL      \undefined \def \showURL       {\relax}        \fi
\providecommand\bibfield[2]{#2}
\providecommand\bibinfo[2]{#2}
\providecommand\natexlab[1]{#1}
\providecommand\showeprint[2][]{arXiv:#2}

\bibitem[\protect\citeauthoryear{Bian, Li, Yang, Chen, and Lin}{Bian
  et~al\mbox{.}}{2017}]%
        {dynamic-clip-17}
\bibfield{author}{\bibinfo{person}{Weijie Bian}, \bibinfo{person}{Si Li},
  \bibinfo{person}{Zhao Yang}, \bibinfo{person}{Guang Chen}, {and}
  \bibinfo{person}{Zhiqing Lin}.} \bibinfo{year}{2017}\natexlab{}.
\newblock \showarticletitle{A Compare-Aggregate Model with Dynamic-Clip
  Attention for Answer Selection}. In \bibinfo{booktitle}{\emph{CIKM}}.
  \bibinfo{pages}{1987--1990}.
\newblock


\bibitem[\protect\citeauthoryear{Bowman, Angeli, Potts, and Manning}{Bowman
  et~al\mbox{.}}{2015}]%
        {DBLP:conf/emnlp/snli}
\bibfield{author}{\bibinfo{person}{Samuel~R. Bowman}, \bibinfo{person}{Gabor
  Angeli}, \bibinfo{person}{Christopher Potts}, {and}
  \bibinfo{person}{Christopher~D. Manning}.} \bibinfo{year}{2015}\natexlab{}.
\newblock \showarticletitle{A large annotated corpus for learning natural
  language inference}. In \bibinfo{booktitle}{\emph{EMNLP}}.
  \bibinfo{pages}{632--642}.
\newblock


\bibitem[\protect\citeauthoryear{Carmel, Lewin{-}Eytan, and Maarek}{Carmel
  et~al\mbox{.}}{2018}]%
        {DBLP:conf/sigir/CarmelLM18}
\bibfield{author}{\bibinfo{person}{David Carmel}, \bibinfo{person}{Liane
  Lewin{-}Eytan}, {and} \bibinfo{person}{Yoelle Maarek}.}
  \bibinfo{year}{2018}\natexlab{}.
\newblock \showarticletitle{Product Question Answering Using Customer Generated
  Content - Research Challenges}. In \bibinfo{booktitle}{\emph{SIGIR}}.
  \bibinfo{pages}{1349--1350}.
\newblock


\bibitem[\protect\citeauthoryear{Chen, Qiu, Yang, Zhou, Huang, Li, and
  Bao}{Chen et~al\mbox{.}}{2019d}]%
        {DBLP:conf/www/ChenQYZHLB19}
\bibfield{author}{\bibinfo{person}{Cen Chen}, \bibinfo{person}{Minghui Qiu},
  \bibinfo{person}{Yinfei Yang}, \bibinfo{person}{Jun Zhou},
  \bibinfo{person}{Jun Huang}, \bibinfo{person}{Xiaolong Li}, {and}
  \bibinfo{person}{Forrest~Sheng Bao}.} \bibinfo{year}{2019}\natexlab{d}.
\newblock \showarticletitle{Multi-Domain Gated {CNN} for Review Helpfulness
  Prediction}. In \bibinfo{booktitle}{\emph{WWW}}. \bibinfo{pages}{2630--2636}.
\newblock


\bibitem[\protect\citeauthoryear{Chen, Guan, Zhao, Zhao, Wang, Zhao, and
  Sun}{Chen et~al\mbox{.}}{2019b}]%
        {DBLP:conf/aaai/0007GZZWZS19}
\bibfield{author}{\bibinfo{person}{Long Chen}, \bibinfo{person}{Ziyu Guan},
  \bibinfo{person}{Wei Zhao}, \bibinfo{person}{Wanqing Zhao},
  \bibinfo{person}{Xiaopeng Wang}, \bibinfo{person}{Zhou Zhao}, {and}
  \bibinfo{person}{Huan Sun}.} \bibinfo{year}{2019}\natexlab{b}.
\newblock \showarticletitle{Answer Identification from Product Reviews for User
  Questions by Multi-Task Attentive Networks}. In
  \bibinfo{booktitle}{\emph{AAAI}}. \bibinfo{pages}{45--52}.
\newblock


\bibitem[\protect\citeauthoryear{Chen, Zhu, Ling, Wei, Jiang, and Inkpen}{Chen
  et~al\mbox{.}}{2017}]%
        {DBLP:conf/acl/esim-17}
\bibfield{author}{\bibinfo{person}{Qian Chen}, \bibinfo{person}{Xiaodan Zhu},
  \bibinfo{person}{Zhen{-}Hua Ling}, \bibinfo{person}{Si Wei},
  \bibinfo{person}{Hui Jiang}, {and} \bibinfo{person}{Diana Inkpen}.}
  \bibinfo{year}{2017}\natexlab{}.
\newblock \showarticletitle{Enhanced {LSTM} for Natural Language Inference}. In
  \bibinfo{booktitle}{\emph{ACL}}. \bibinfo{pages}{1657--1668}.
\newblock


\bibitem[\protect\citeauthoryear{Chen, Li, Ji, Zhou, and Chen}{Chen
  et~al\mbox{.}}{2019c}]%
        {DBLP:conf/wsdm/ChenLJZC19}
\bibfield{author}{\bibinfo{person}{Shiqian Chen}, \bibinfo{person}{Chenliang
  Li}, \bibinfo{person}{Feng Ji}, \bibinfo{person}{Wei Zhou}, {and}
  \bibinfo{person}{Haiqing Chen}.} \bibinfo{year}{2019}\natexlab{c}.
\newblock \showarticletitle{Review-Driven Answer Generation for Product-Related
  Questions in E-Commerce}. In \bibinfo{booktitle}{\emph{WSDM}}.
  \bibinfo{pages}{411--419}.
\newblock


\bibitem[\protect\citeauthoryear{Chen, Gu, Ren, He, Xie, Guo, Yin, and
  Zhang}{Chen et~al\mbox{.}}{2019a}]%
        {DBLP:conf/ijcai/ChenGRHXGYZ19}
\bibfield{author}{\bibinfo{person}{Weijian Chen}, \bibinfo{person}{Yulong Gu},
  \bibinfo{person}{Zhaochun Ren}, \bibinfo{person}{Xiangnan He},
  \bibinfo{person}{Hongtao Xie}, \bibinfo{person}{Tong Guo},
  \bibinfo{person}{Dawei Yin}, {and} \bibinfo{person}{Yongdong Zhang}.}
  \bibinfo{year}{2019}\natexlab{a}.
\newblock \showarticletitle{Semi-supervised User Profiling with Heterogeneous
  Graph Attention Networks}. In \bibinfo{booktitle}{\emph{IJCAI}}.
  \bibinfo{pages}{2116--2122}.
\newblock


\bibitem[\protect\citeauthoryear{Deng, Lam, Xie, Chen, Li, Yang, and Shen}{Deng
  et~al\mbox{.}}{2019}]%
        {aaai20-dy}
\bibfield{author}{\bibinfo{person}{Yang Deng}, \bibinfo{person}{Wai Lam},
  \bibinfo{person}{Yuexiang Xie}, \bibinfo{person}{Daoyuan Chen},
  \bibinfo{person}{Yaliang Li}, \bibinfo{person}{Min Yang}, {and}
  \bibinfo{person}{Ying Shen}.} \bibinfo{year}{2019}\natexlab{}.
\newblock \showarticletitle{Joint Learning of Answer Selection and Answer
  Summary Generation in Community Question Answering}.
\newblock \bibinfo{journal}{\emph{CoRR}}  \bibinfo{volume}{abs/1911.09801}
  (\bibinfo{year}{2019}).
\newblock


\bibitem[\protect\citeauthoryear{Deng, Shen, Yang, Li, Du, Fan, and Lei}{Deng
  et~al\mbox{.}}{2018}]%
        {DBLP:conf/coling/DengSYLDFL18}
\bibfield{author}{\bibinfo{person}{Yang Deng}, \bibinfo{person}{Ying Shen},
  \bibinfo{person}{Min Yang}, \bibinfo{person}{Yaliang Li},
  \bibinfo{person}{Nan Du}, \bibinfo{person}{Wei Fan}, {and}
  \bibinfo{person}{Kai Lei}.} \bibinfo{year}{2018}\natexlab{}.
\newblock \showarticletitle{Knowledge as {A} Bridge: Improving Cross-domain
  Answer Selection with External Knowledge}. In
  \bibinfo{booktitle}{\emph{COLING}}. \bibinfo{pages}{3295--3305}.
\newblock


\bibitem[\protect\citeauthoryear{Eric, Krishnan, Charette, and Manning}{Eric
  et~al\mbox{.}}{2017}]%
        {DBLP:conf/sigdial/EricKCM17}
\bibfield{author}{\bibinfo{person}{Mihail Eric}, \bibinfo{person}{Lakshmi
  Krishnan}, \bibinfo{person}{Francois Charette}, {and}
  \bibinfo{person}{Christopher~D. Manning}.} \bibinfo{year}{2017}\natexlab{}.
\newblock \showarticletitle{Key-Value Retrieval Networks for Task-Oriented
  Dialogue}. In \bibinfo{booktitle}{\emph{SIGdial}}. \bibinfo{pages}{37--49}.
\newblock


\bibitem[\protect\citeauthoryear{Fan, Feng, Guo, Sun, and Li}{Fan
  et~al\mbox{.}}{2019}]%
        {DBLP:conf/www/FanFGSL19}
\bibfield{author}{\bibinfo{person}{Miao Fan}, \bibinfo{person}{Chao Feng},
  \bibinfo{person}{Lin Guo}, \bibinfo{person}{Mingming Sun}, {and}
  \bibinfo{person}{Ping Li}.} \bibinfo{year}{2019}\natexlab{}.
\newblock \showarticletitle{Product-Aware Helpfulness Prediction of Online
  Reviews}. In \bibinfo{booktitle}{\emph{WWW}}. \bibinfo{pages}{2715--2721}.
\newblock


\bibitem[\protect\citeauthoryear{Gao, Ren, Zhao, Zhao, Yin, and Yan}{Gao
  et~al\mbox{.}}{2019}]%
        {DBLP:conf/wsdm/GaoRZZYY19}
\bibfield{author}{\bibinfo{person}{Shen Gao}, \bibinfo{person}{Zhaochun Ren},
  \bibinfo{person}{Yihong Zhao}, \bibinfo{person}{Dongyan Zhao},
  \bibinfo{person}{Dawei Yin}, {and} \bibinfo{person}{Rui Yan}.}
  \bibinfo{year}{2019}\natexlab{}.
\newblock \showarticletitle{Product-Aware Answer Generation in E-Commerce
  Question-Answering}. In \bibinfo{booktitle}{\emph{WSDM}}.
  \bibinfo{pages}{429--437}.
\newblock


\bibitem[\protect\citeauthoryear{Glorot and Bengio}{Glorot and Bengio}{2010}]%
        {DBLP:journals/jmlr/xavier}
\bibfield{author}{\bibinfo{person}{Xavier Glorot} {and} \bibinfo{person}{Yoshua
  Bengio}.} \bibinfo{year}{2010}\natexlab{}.
\newblock \showarticletitle{Understanding the difficulty of training deep
  feedforward neural networks}. In \bibinfo{booktitle}{\emph{AISTATS}}.
  \bibinfo{pages}{249--256}.
\newblock


\bibitem[\protect\citeauthoryear{Halder, Kan, and Sugiyama}{Halder
  et~al\mbox{.}}{2019}]%
        {DBLP:conf/naacl/HalderKS19}
\bibfield{author}{\bibinfo{person}{Kishaloy Halder}, \bibinfo{person}{Min{-}Yen
  Kan}, {and} \bibinfo{person}{Kazunari Sugiyama}.}
  \bibinfo{year}{2019}\natexlab{}.
\newblock \showarticletitle{Predicting Helpful Posts in Open-Ended Discussion
  Forums: {A} Neural Architecture}. In \bibinfo{booktitle}{\emph{NAACL-HLT}}.
  \bibinfo{pages}{3148--3157}.
\newblock


\bibitem[\protect\citeauthoryear{Harabagiu and Hickl}{Harabagiu and
  Hickl}{2006}]%
        {DBLP:conf/acl/HarabagiuH06}
\bibfield{author}{\bibinfo{person}{Sanda~M. Harabagiu} {and}
  \bibinfo{person}{Andrew Hickl}.} \bibinfo{year}{2006}\natexlab{}.
\newblock \showarticletitle{Methods for Using Textual Entailment in Open-Domain
  Question Answering}. In \bibinfo{booktitle}{\emph{ACL}}.
\newblock


\bibitem[\protect\citeauthoryear{Harel, Albo, Agichtein, and Radinsky}{Harel
  et~al\mbox{.}}{2019}]%
        {DBLP:conf/www/HarelAAR19}
\bibfield{author}{\bibinfo{person}{Shahar Harel}, \bibinfo{person}{Sefi Albo},
  \bibinfo{person}{Eugene Agichtein}, {and} \bibinfo{person}{Kira Radinsky}.}
  \bibinfo{year}{2019}\natexlab{}.
\newblock \showarticletitle{Learning Novelty-Aware Ranking of Answers to
  Complex Questions}. In \bibinfo{booktitle}{\emph{WWW}}.
  \bibinfo{pages}{2799--2805}.
\newblock


\bibitem[\protect\citeauthoryear{He and McAuley}{He and McAuley}{2016}]%
        {DBLP:conf/www/amazon-review-16}
\bibfield{author}{\bibinfo{person}{Ruining He} {and} \bibinfo{person}{Julian~J.
  McAuley}.} \bibinfo{year}{2016}\natexlab{}.
\newblock \showarticletitle{Ups and Downs: Modeling the Visual Evolution of
  Fashion Trends with One-Class Collaborative Filtering}. In
  \bibinfo{booktitle}{\emph{WWW}}. \bibinfo{pages}{507--517}.
\newblock


\bibitem[\protect\citeauthoryear{Jeon, Croft, Lee, and Park}{Jeon
  et~al\mbox{.}}{2006}]%
        {DBLP:conf/sigir/answer-unreliable-2}
\bibfield{author}{\bibinfo{person}{Jiwoon Jeon}, \bibinfo{person}{W.~Bruce
  Croft}, \bibinfo{person}{Joon~Ho Lee}, {and} \bibinfo{person}{Soyeon Park}.}
  \bibinfo{year}{2006}\natexlab{}.
\newblock \showarticletitle{A framework to predict the quality of answers with
  non-textual features}. In \bibinfo{booktitle}{\emph{SIGIR}}.
  \bibinfo{pages}{228--235}.
\newblock


\bibitem[\protect\citeauthoryear{Kingma and Ba}{Kingma and Ba}{2014}]%
        {kingma2014adam}
\bibfield{author}{\bibinfo{person}{Diederik~P Kingma} {and}
  \bibinfo{person}{Jimmy Ba}.} \bibinfo{year}{2014}\natexlab{}.
\newblock \showarticletitle{Adam: A method for stochastic optimization}.
\newblock \bibinfo{journal}{\emph{arXiv preprint arXiv:1412.6980}}
  (\bibinfo{year}{2014}).
\newblock


\bibitem[\protect\citeauthoryear{Kipf and Welling}{Kipf and Welling}{2017}]%
        {DBLP:conf/iclr/gcn-17}
\bibfield{author}{\bibinfo{person}{Thomas~N. Kipf} {and} \bibinfo{person}{Max
  Welling}.} \bibinfo{year}{2017}\natexlab{}.
\newblock \showarticletitle{Semi-Supervised Classification with Graph
  Convolutional Networks}. In \bibinfo{booktitle}{\emph{ICLR}}.
\newblock


\bibitem[\protect\citeauthoryear{Li, Qin, Liu, Yang, and Li}{Li
  et~al\mbox{.}}{2019}]%
        {DBLP:conf/cikm/LiQLYL19}
\bibfield{author}{\bibinfo{person}{Ao Li}, \bibinfo{person}{Zhou Qin},
  \bibinfo{person}{Runshi Liu}, \bibinfo{person}{Yiqun Yang}, {and}
  \bibinfo{person}{Dong Li}.} \bibinfo{year}{2019}\natexlab{}.
\newblock \showarticletitle{Spam Review Detection with Graph Convolutional
  Networks}. In \bibinfo{booktitle}{\emph{CIKM}}. \bibinfo{pages}{2703--2711}.
\newblock


\bibitem[\protect\citeauthoryear{Liu}{Liu}{2009}]%
        {DBLP:journals/ftir/Liu09}
\bibfield{author}{\bibinfo{person}{Tie{-}Yan Liu}.}
  \bibinfo{year}{2009}\natexlab{}.
\newblock \showarticletitle{Learning to Rank for Information Retrieval}.
\newblock \bibinfo{journal}{\emph{Foundations and Trends in Information
  Retrieval}} \bibinfo{volume}{3}, \bibinfo{number}{3} (\bibinfo{year}{2009}),
  \bibinfo{pages}{225--331}.
\newblock


\bibitem[\protect\citeauthoryear{Lyu, Ouyang, Wang, Shen, and Cheng}{Lyu
  et~al\mbox{.}}{2019}]%
        {DBLP:conf/www/LyuOWSC19}
\bibfield{author}{\bibinfo{person}{Shanshan Lyu}, \bibinfo{person}{Wentao
  Ouyang}, \bibinfo{person}{Yongqing Wang}, \bibinfo{person}{Huawei Shen},
  {and} \bibinfo{person}{Xueqi Cheng}.} \bibinfo{year}{2019}\natexlab{}.
\newblock \showarticletitle{What We Vote for? Answer Selection from User
  Expertise View in Community Question Answering}. In
  \bibinfo{booktitle}{\emph{The World Wide Web Conference, {WWW} 2019, San
  Francisco, CA, USA, May 13-17, 2019}}. \bibinfo{pages}{1198--1209}.
\newblock


\bibitem[\protect\citeauthoryear{McAuley and Yang}{McAuley and Yang}{2016}]%
        {DBLP:conf/www/McAuleyY16}
\bibfield{author}{\bibinfo{person}{Julian McAuley} {and} \bibinfo{person}{Alex
  Yang}.} \bibinfo{year}{2016}\natexlab{}.
\newblock \showarticletitle{Addressing Complex and Subjective Product-Related
  Queries with Customer Reviews}. In \bibinfo{booktitle}{\emph{WWW}}.
  \bibinfo{pages}{625--635}.
\newblock


\bibitem[\protect\citeauthoryear{Nakov, Hoogeveen, M{\`{a}}rquez, Moschitti,
  Mubarak, Baldwin, and Verspoor}{Nakov et~al\mbox{.}}{2017}]%
        {DBLP:conf/semeval/NakovHMMMBV17}
\bibfield{author}{\bibinfo{person}{Preslav Nakov}, \bibinfo{person}{Doris
  Hoogeveen}, \bibinfo{person}{Llu{\'{\i}}s M{\`{a}}rquez},
  \bibinfo{person}{Alessandro Moschitti}, \bibinfo{person}{Hamdy Mubarak},
  \bibinfo{person}{Timothy Baldwin}, {and} \bibinfo{person}{Karin Verspoor}.}
  \bibinfo{year}{2017}\natexlab{}.
\newblock \showarticletitle{SemEval-2017 Task 3: Community Question Answering}.
  In \bibinfo{booktitle}{\emph{SemEval@ACL}}. \bibinfo{pages}{27--48}.
\newblock


\bibitem[\protect\citeauthoryear{Ni, Li, and McAuley}{Ni et~al\mbox{.}}{2019}]%
        {DBLP:conf/emnlp/NiLM19}
\bibfield{author}{\bibinfo{person}{Jianmo Ni}, \bibinfo{person}{Jiacheng Li},
  {and} \bibinfo{person}{Julian~J. McAuley}.} \bibinfo{year}{2019}\natexlab{}.
\newblock \showarticletitle{Justifying Recommendations using Distantly-Labeled
  Reviews and Fine-Grained Aspects}. In
  \bibinfo{booktitle}{\emph{EMNLP-IJCNLP}}. \bibinfo{pages}{188--197}.
\newblock


\bibitem[\protect\citeauthoryear{Omari, Carmel, Rokhlenko, and Szpektor}{Omari
  et~al\mbox{.}}{2016}]%
        {DBLP:conf/sigir/OmariCRS16}
\bibfield{author}{\bibinfo{person}{Adi Omari}, \bibinfo{person}{David Carmel},
  \bibinfo{person}{Oleg Rokhlenko}, {and} \bibinfo{person}{Idan Szpektor}.}
  \bibinfo{year}{2016}\natexlab{}.
\newblock \showarticletitle{Novelty based Ranking of Human Answers for
  Community Questions}. In \bibinfo{booktitle}{\emph{SIGIR}}.
  \bibinfo{pages}{215--224}.
\newblock


\bibitem[\protect\citeauthoryear{Qiu and Huang}{Qiu and Huang}{2015}]%
        {DBLP:conf/ijcai/Qiu-yahooqa-15}
\bibfield{author}{\bibinfo{person}{Xipeng Qiu} {and} \bibinfo{person}{Xuanjing
  Huang}.} \bibinfo{year}{2015}\natexlab{}.
\newblock \showarticletitle{Convolutional Neural Tensor Network Architecture
  for Community-Based Question Answering}. In
  \bibinfo{booktitle}{\emph{IJCAI}}. \bibinfo{pages}{1305--1311}.
\newblock


\bibitem[\protect\citeauthoryear{Rao, He, and Lin}{Rao et~al\mbox{.}}{2017}]%
        {DBLP:conf/sigir/RaoHL17}
\bibfield{author}{\bibinfo{person}{Jinfeng Rao}, \bibinfo{person}{Hua He},
  {and} \bibinfo{person}{Jimmy Lin}.} \bibinfo{year}{2017}\natexlab{}.
\newblock \showarticletitle{Experiments with Convolutional Neural Network
  Models for Answer Selection}. In \bibinfo{booktitle}{\emph{SIGIR}}.
  \bibinfo{pages}{1217--1220}.
\newblock


\bibitem[\protect\citeauthoryear{Rao, Liu, Tay, Yang, Shi, and Lin}{Rao
  et~al\mbox{.}}{2019}]%
        {rao2019bridging}
\bibfield{author}{\bibinfo{person}{Jinfeng Rao}, \bibinfo{person}{Linqing Liu},
  \bibinfo{person}{Yi Tay}, \bibinfo{person}{Wei Yang}, \bibinfo{person}{Peng
  Shi}, {and} \bibinfo{person}{Jimmy Lin}.} \bibinfo{year}{2019}\natexlab{}.
\newblock \showarticletitle{Bridging the Gap Between Relevance Matching and
  Semantic Matching for Short Text Similarity Modeling}. In
  \bibinfo{booktitle}{\emph{EMNLP-IJCNLP}}. \bibinfo{pages}{5373--5384}.
\newblock


\bibitem[\protect\citeauthoryear{Robertson and Zaragoza}{Robertson and
  Zaragoza}{2009}]%
        {DBLP:journals/ftir/bm25}
\bibfield{author}{\bibinfo{person}{Stephen~E. Robertson} {and}
  \bibinfo{person}{Hugo Zaragoza}.} \bibinfo{year}{2009}\natexlab{}.
\newblock \showarticletitle{The Probabilistic Relevance Framework: {BM25} and
  Beyond}.
\newblock \bibinfo{journal}{\emph{Foundations and Trends in Information
  Retrieval}} \bibinfo{volume}{3}, \bibinfo{number}{4} (\bibinfo{year}{2009}),
  \bibinfo{pages}{333--389}.
\newblock


\bibitem[\protect\citeauthoryear{Schlichtkrull, Kipf, Bloem, van~den Berg,
  Titov, and Welling}{Schlichtkrull et~al\mbox{.}}{2018}]%
        {DBLP:conf/esws/relation-gcn-18}
\bibfield{author}{\bibinfo{person}{Michael~Sejr Schlichtkrull},
  \bibinfo{person}{Thomas~N. Kipf}, \bibinfo{person}{Peter Bloem},
  \bibinfo{person}{Rianne van~den Berg}, \bibinfo{person}{Ivan Titov}, {and}
  \bibinfo{person}{Max Welling}.} \bibinfo{year}{2018}\natexlab{}.
\newblock \showarticletitle{Modeling Relational Data with Graph Convolutional
  Networks}. In \bibinfo{booktitle}{\emph{ESWC}}. \bibinfo{pages}{593--607}.
\newblock


\bibitem[\protect\citeauthoryear{Severyn and Moschitti}{Severyn and
  Moschitti}{2013}]%
        {DBLP:conf/emnlp/SeverynM13}
\bibfield{author}{\bibinfo{person}{Aliaksei Severyn} {and}
  \bibinfo{person}{Alessandro Moschitti}.} \bibinfo{year}{2013}\natexlab{}.
\newblock \showarticletitle{Automatic Feature Engineering for Answer Selection
  and Extraction}. In \bibinfo{booktitle}{\emph{EMNLP}}.
  \bibinfo{pages}{458--467}.
\newblock


\bibitem[\protect\citeauthoryear{Severyn and Moschitti}{Severyn and
  Moschitti}{2015}]%
        {DBLP:conf/sigir/SeverynM15}
\bibfield{author}{\bibinfo{person}{Aliaksei Severyn} {and}
  \bibinfo{person}{Alessandro Moschitti}.} \bibinfo{year}{2015}\natexlab{}.
\newblock \showarticletitle{Learning to Rank Short Text Pairs with
  Convolutional Deep Neural Networks}. In \bibinfo{booktitle}{\emph{SIGIR}}.
  \bibinfo{pages}{373--382}.
\newblock


\bibitem[\protect\citeauthoryear{Shah and Pomerantz}{Shah and
  Pomerantz}{2010}]%
        {DBLP:conf/sigir/answer-quality-2010}
\bibfield{author}{\bibinfo{person}{Chirag Shah} {and} \bibinfo{person}{Jeffrey
  Pomerantz}.} \bibinfo{year}{2010}\natexlab{}.
\newblock \showarticletitle{Evaluating and predicting answer quality in
  community {QA}}. In \bibinfo{booktitle}{\emph{SIGIR}}.
  \bibinfo{pages}{411--418}.
\newblock


\bibitem[\protect\citeauthoryear{Shao, Cai, Chen, and de~Rijke}{Shao
  et~al\mbox{.}}{2019}]%
        {DBLP:conf/sigir/ShaoCCR19}
\bibfield{author}{\bibinfo{person}{Taihua Shao}, \bibinfo{person}{Fei Cai},
  \bibinfo{person}{Honghui Chen}, {and} \bibinfo{person}{Maarten de Rijke}.}
  \bibinfo{year}{2019}\natexlab{}.
\newblock \showarticletitle{Length-adaptive Neural Network for Answer
  Selection}. In \bibinfo{booktitle}{\emph{SIGIR}}. \bibinfo{pages}{869--872}.
\newblock


\bibitem[\protect\citeauthoryear{Sun, Tang, Du, Deng, and Nie}{Sun
  et~al\mbox{.}}{2019}]%
        {DBLP:conf/sigir/SunTDDN19}
\bibfield{author}{\bibinfo{person}{Zhiqing Sun}, \bibinfo{person}{Jian Tang},
  \bibinfo{person}{Pan Du}, \bibinfo{person}{Zhi{-}Hong Deng}, {and}
  \bibinfo{person}{Jian{-}Yun Nie}.} \bibinfo{year}{2019}\natexlab{}.
\newblock \showarticletitle{DivGraphPointer: {A} Graph Pointer Network for
  Extracting Diverse Keyphrases}. In \bibinfo{booktitle}{\emph{SIGIR}}.
  \bibinfo{pages}{755--764}.
\newblock


\bibitem[\protect\citeauthoryear{Suryanto, Lim, Sun, and Chiang}{Suryanto
  et~al\mbox{.}}{2009}]%
        {DBLP:conf/wsdm/answer-unreliable}
\bibfield{author}{\bibinfo{person}{Maggy~Anastasia Suryanto},
  \bibinfo{person}{Ee{-}Peng Lim}, \bibinfo{person}{Aixin Sun}, {and}
  \bibinfo{person}{Roger H.~L. Chiang}.} \bibinfo{year}{2009}\natexlab{}.
\newblock \showarticletitle{Quality-aware collaborative question answering:
  methods and evaluation}. In \bibinfo{booktitle}{\emph{WSDM}}.
  \bibinfo{pages}{142--151}.
\newblock


\bibitem[\protect\citeauthoryear{Tan, dos Santos, Xiang, and Zhou}{Tan
  et~al\mbox{.}}{2016}]%
        {DBLP:conf/acl/TanSXZ16}
\bibfield{author}{\bibinfo{person}{Ming Tan},
  \bibinfo{person}{C{\'{\i}}cero~Nogueira dos Santos}, \bibinfo{person}{Bing
  Xiang}, {and} \bibinfo{person}{Bowen Zhou}.} \bibinfo{year}{2016}\natexlab{}.
\newblock \showarticletitle{Improved Representation Learning for Question
  Answer Matching}. In \bibinfo{booktitle}{\emph{ACL}}.
\newblock


\bibitem[\protect\citeauthoryear{Tay, Phan, Luu, and Hui}{Tay
  et~al\mbox{.}}{2017}]%
        {DBLP:conf/sigir/yitay-yahooqa}
\bibfield{author}{\bibinfo{person}{Yi Tay}, \bibinfo{person}{Minh~C. Phan},
  \bibinfo{person}{Anh~Tuan Luu}, {and} \bibinfo{person}{Siu~Cheung Hui}.}
  \bibinfo{year}{2017}\natexlab{}.
\newblock \showarticletitle{Learning to Rank Question Answer Pairs with
  Holographic Dual {LSTM} Architecture}. In \bibinfo{booktitle}{\emph{SIGIR}}.
  \bibinfo{pages}{695--704}.
\newblock


\bibitem[\protect\citeauthoryear{Trivedi, Kwon, Khot, Sabharwal, and
  Balasubramanian}{Trivedi et~al\mbox{.}}{2019}]%
        {DBLP:conf/naacl/TrivediKKSB19}
\bibfield{author}{\bibinfo{person}{Harsh Trivedi}, \bibinfo{person}{Heeyoung
  Kwon}, \bibinfo{person}{Tushar Khot}, \bibinfo{person}{Ashish Sabharwal},
  {and} \bibinfo{person}{Niranjan Balasubramanian}.}
  \bibinfo{year}{2019}\natexlab{}.
\newblock \showarticletitle{Repurposing Entailment for Multi-Hop Question
  Answering Tasks}. In \bibinfo{booktitle}{\emph{NAACL-HLT}}.
  \bibinfo{pages}{2948--2958}.
\newblock


\bibitem[\protect\citeauthoryear{Tymoshenko and Moschitti}{Tymoshenko and
  Moschitti}{2018}]%
        {DBLP:conf/emnlp/TymoshenkoM18}
\bibfield{author}{\bibinfo{person}{Kateryna Tymoshenko} {and}
  \bibinfo{person}{Alessandro Moschitti}.} \bibinfo{year}{2018}\natexlab{}.
\newblock \showarticletitle{Cross-Pair Text Representations for Answer Sentence
  Selection}. In \bibinfo{booktitle}{\emph{NAACL}}.
  \bibinfo{pages}{2162--2173}.
\newblock


\bibitem[\protect\citeauthoryear{Wan and McAuley}{Wan and McAuley}{2016}]%
        {DBLP:conf/icdm/WanM16}
\bibfield{author}{\bibinfo{person}{Mengting Wan} {and}
  \bibinfo{person}{Julian~J. McAuley}.} \bibinfo{year}{2016}\natexlab{}.
\newblock \showarticletitle{Modeling Ambiguity, Subjectivity, and Diverging
  Viewpoints in Opinion Question Answering Systems}. In
  \bibinfo{booktitle}{\emph{ICDM}}. \bibinfo{pages}{489--498}.
\newblock


\bibitem[\protect\citeauthoryear{Wang, Smith, and Mitamura}{Wang
  et~al\mbox{.}}{2007}]%
        {DBLP:conf/emnlp/WangSM07}
\bibfield{author}{\bibinfo{person}{Mengqiu Wang}, \bibinfo{person}{Noah~A.
  Smith}, {and} \bibinfo{person}{Teruko Mitamura}.}
  \bibinfo{year}{2007}\natexlab{}.
\newblock \showarticletitle{What is the Jeopardy Model? {A} Quasi-Synchronous
  Grammar for {QA}}. In \bibinfo{booktitle}{\emph{EMNLP-CoNLL}}.
  \bibinfo{pages}{22--32}.
\newblock


\bibitem[\protect\citeauthoryear{Wang and Jiang}{Wang and Jiang}{2017}]%
        {DBLP:conf/iclr/Wang017}
\bibfield{author}{\bibinfo{person}{Shuohang Wang} {and} \bibinfo{person}{Jing
  Jiang}.} \bibinfo{year}{2017}\natexlab{}.
\newblock \showarticletitle{A Compare-Aggregate Model for Matching Text
  Sequences}. In \bibinfo{booktitle}{\emph{ICLR}}.
\newblock


\bibitem[\protect\citeauthoryear{Wang, Hamza, and Florian}{Wang
  et~al\mbox{.}}{2017}]%
        {DBLP:conf/ijcai/WangHF17}
\bibfield{author}{\bibinfo{person}{Zhiguo Wang}, \bibinfo{person}{Wael Hamza},
  {and} \bibinfo{person}{Radu Florian}.} \bibinfo{year}{2017}\natexlab{}.
\newblock \showarticletitle{Bilateral Multi-Perspective Matching for Natural
  Language Sentences}. In \bibinfo{booktitle}{\emph{IJCAI}}.
  \bibinfo{pages}{4144--4150}.
\newblock


\bibitem[\protect\citeauthoryear{Yang, Ai, Guo, and Croft}{Yang
  et~al\mbox{.}}{2016}]%
        {DBLP:conf/cikm/anmm}
\bibfield{author}{\bibinfo{person}{Liu Yang}, \bibinfo{person}{Qingyao Ai},
  \bibinfo{person}{Jiafeng Guo}, {and} \bibinfo{person}{W.~Bruce Croft}.}
  \bibinfo{year}{2016}\natexlab{}.
\newblock \showarticletitle{aNMM: Ranking Short Answer Texts with
  Attention-Based Neural Matching Model}. In \bibinfo{booktitle}{\emph{CIKM}}.
  \bibinfo{pages}{287--296}.
\newblock


\bibitem[\protect\citeauthoryear{Yang, Zhang, Gao, Ji, and Chen}{Yang
  et~al\mbox{.}}{2019}]%
        {DBLP:conf/acl/re2-2019}
\bibfield{author}{\bibinfo{person}{Runqi Yang}, \bibinfo{person}{Jianhai
  Zhang}, \bibinfo{person}{Xing Gao}, \bibinfo{person}{Feng Ji}, {and}
  \bibinfo{person}{Haiqing Chen}.} \bibinfo{year}{2019}\natexlab{}.
\newblock \showarticletitle{Simple and Effective Text Matching with Richer
  Alignment Features}. In \bibinfo{booktitle}{\emph{ACL}}.
  \bibinfo{pages}{4699--4709}.
\newblock


\bibitem[\protect\citeauthoryear{Yang, Yih, and Meek}{Yang
  et~al\mbox{.}}{2015}]%
        {DBLP:conf/emnlp/wikiqa}
\bibfield{author}{\bibinfo{person}{Yi Yang}, \bibinfo{person}{Wen{-}tau Yih},
  {and} \bibinfo{person}{Christopher Meek}.} \bibinfo{year}{2015}\natexlab{}.
\newblock \showarticletitle{WikiQA: {A} Challenge Dataset for Open-Domain
  Question Answering}. In \bibinfo{booktitle}{\emph{EMNLP}}.
  \bibinfo{pages}{2013--2018}.
\newblock


\bibitem[\protect\citeauthoryear{Yu, Qiu, Jiang, Huang, Song, Chu, and Chen}{Yu
  et~al\mbox{.}}{2018}]%
        {DBLP:conf/wsdm/YuQJHSCC18}
\bibfield{author}{\bibinfo{person}{Jianfei Yu}, \bibinfo{person}{Minghui Qiu},
  \bibinfo{person}{Jing Jiang}, \bibinfo{person}{Jun Huang},
  \bibinfo{person}{Shuangyong Song}, \bibinfo{person}{Wei Chu}, {and}
  \bibinfo{person}{Haiqing Chen}.} \bibinfo{year}{2018}\natexlab{}.
\newblock \showarticletitle{Modelling Domain Relationships for Transfer
  Learning on Retrieval-based Question Answering Systems in E-commerce}. In
  \bibinfo{booktitle}{\emph{WSDM}}. \bibinfo{pages}{682--690}.
\newblock


\bibitem[\protect\citeauthoryear{Yu, Zha, and Chua}{Yu et~al\mbox{.}}{2012}]%
        {DBLP:conf/emnlp/YuZC12}
\bibfield{author}{\bibinfo{person}{Jianxing Yu}, \bibinfo{person}{Zheng{-}Jun
  Zha}, {and} \bibinfo{person}{Tat{-}Seng Chua}.}
  \bibinfo{year}{2012}\natexlab{}.
\newblock \showarticletitle{Answering Opinion Questions on Products by
  Exploiting Hierarchical Organization of Consumer Reviews}. In
  \bibinfo{booktitle}{\emph{EMNLP-CoNLL}}. \bibinfo{pages}{391--401}.
\newblock


\bibitem[\protect\citeauthoryear{Yu and Lam}{Yu and Lam}{2018}]%
        {DBLP:conf/wsdm/YuL18}
\bibfield{author}{\bibinfo{person}{Qian Yu} {and} \bibinfo{person}{Wai Lam}.}
  \bibinfo{year}{2018}\natexlab{}.
\newblock \showarticletitle{Review-Aware Answer Prediction for Product-Related
  Questions Incorporating Aspects}. In \bibinfo{booktitle}{\emph{WSDM}}.
  \bibinfo{pages}{691--699}.
\newblock


\bibitem[\protect\citeauthoryear{Zhang, Lam, Deng, and Ma}{Zhang
  et~al\mbox{.}}{2020}]%
        {Zhang-WWW2020-rahp}
\bibfield{author}{\bibinfo{person}{Wenxuan Zhang}, \bibinfo{person}{Wai Lam},
  \bibinfo{person}{Yang Deng}, {and} \bibinfo{person}{Jing Ma}.}
  \bibinfo{year}{2020}\natexlab{}.
\newblock \showarticletitle{Review-guided Helpful Answer Identification in
  E-Commerce}. In \bibinfo{booktitle}{\emph{WWW}}.
  \bibinfo{pages}{2620–2626}.
\newblock


\end{thebibliography}

\end{document}